\documentclass[sigconf,screen]{acmart}

\usepackage{soul}
\usepackage{multirow}
\usepackage{epstopdf}
\usepackage{hyperref}
\usepackage{listings}
\usepackage{fancybox}
\usepackage{graphicx}
\usepackage{subcaption}
\usepackage{paralist}
\usepackage{ragged2e}
\usepackage{enumitem}
\usepackage{color}
\usepackage{xcolor,colortbl}
\usepackage{balance}

\usepackage[framemethod=TikZ]{mdframed}
\usepackage{tcolorbox}% http://ctan.org/pkg/tcolorbox
\makeatletter
\newcommand{\mybox}[1]{%
  \setbox0=\hbox{#1}%
  \setlength{\@tempdima}{\dimexpr\wd0+13pt}%
  \begin{tcolorbox}[boxrule=0.5pt, colback=white, arc=4pt,
      left=6pt,right=6pt,top=6pt,bottom=6pt,boxsep=0pt]
    #1
  \end{tcolorbox}
}
\usepackage{tikz}
\usepackage{pgfplots}
\usetikzlibrary{pgfplots.statistics,calc}
\usepackage{color}
\usepackage{xcolor}
\usepackage{array}
\usepackage{amsmath}
\usepackage{centernot}
\usepackage{xspace}
\usepackage{url}
\usepackage{verbatim}
\usepackage{wrapfig}
\usepackage{tabularx}
\usepackage{bbding}
\usepackage{pifont}
\usepackage{wasysym}
\usepackage{makecell}

\makeatletter  
\newif\if@restonecol  
\makeatother  
\usepackage[linesnumbered,ruled,vlined]{algorithm2e}%[ruled,vlined]{  
\usepackage{algpseudocode}  
\AtBeginDocument{%
  \providecommand\BibTeX{{%
    \normalfont B\kern-0.5em{\scshape i\kern-0.25em b}\kern-0.8em\TeX}}}

\balance
\setcopyright{rightsretained}
\acmPrice{}
\acmDOI{10.1145/3540250.3549089}
\acmYear{2022}
\copyrightyear{2022}
\acmSubmissionID{fse22main-p84-p}
\acmISBN{978-1-4503-9413-0/22/11}
\acmConference[ESEC/FSE '22]{Proceedings of the 30th ACM Joint European Software Engineering Conference and Symposium on the Foundations of Software Engineering}{November 14--18, 2022}{Singapore, Singapore}
\acmBooktitle{Proceedings of the 30th ACM Joint European Software Engineering Conference and Symposium on the Foundations of Software Engineering (ESEC/FSE '22), November 14--18, 2022, Singapore, Singapore}

\newcommand{\tool}{\texttt{Tscope}}

\newcommand{\extraction}{Tscope}

\begin{document}

%\title{Putting Test Cases under Microscope: A Fine-grained Approach for Test Suite Minimization in Natural Language}
%Putting Test Cases under Microscope: A Fine-grained Approach for Redundancy Detecting in Natural Language
%Putting them under Microscope: A fine-grained Approach for Detecting Redundant Test Cases in Natural Language
%TC_Microscope: A Fine-grained Approach for Detecting Redundant Test Cases in Natural Language
%Putting Test Cases under Microscope: A Fine-grained Approach for Detecting Redundant Test Cases in Natural Language
%Putting Test Cases under Microscope: A Fine-grained Approach for Redundancy Detection in System-level Testing
\title{Putting Them under Microscope: A Fine-Grained Approach for Detecting Redundant Test Cases in Natural Language}

\author{Zhiyuan Chang}
\authornote{Both authors contributed equally to this research.}
\author{Mingyang Li}
\authornotemark[1]
\email{{zhiyuan2019,mingyang2017}@iscas.ac.cn}
\affiliation{%
  \institution{Laboratory for Internet Software Technologies,}
  \department{Institute of Software Chinese Academy of Sciences
  \city{Beijing}
  \country{China}}
}
\affiliation{%
  \institution{
  University of Chinese Academy of Sciences
  \city{Beijing}
  \country{China}}
}

\author{Junjie Wang}
\email{junjie@iscas.ac.cn}
\authornote{Corresponding authors.}
\affiliation{%
  \institution{Laboratory for Internet Software Technologies,}
  \department{Institute of Software Chinese Academy of Sciences
  \city{Beijing}
  \country{China}}
}
\affiliation{%
  \institution{
  University of Chinese Academy of Sciences
  \city{Beijing}
  \country{China}}
}

\author{Qing Wang}
\authornotemark[2]
\email{wq@iscas.ac.cn}
\affiliation{%
  \institution{Laboratory for Internet Software Technologies,}
  \department{Institute of Software Chinese Academy of Sciences
  \city{Beijing}
  \country{China}}
}
\affiliation{%
  \institution{
  State Key Laboratory of Computer Science,}
  \department{Institute of Software Chinese Academy of Sciences
  \city{Beijing}
  \country{China}}
}
\affiliation{%
  \institution{
  University of Chinese Academy of Sciences
  \city{Beijing}
  \country{China}}
}

\author{Shoubin Li}
\email{shoubin@iscas.ac.cn}
\affiliation{%
  \institution{Laboratory for Internet Software Technologies,}
  \department{Institute of Software Chinese Academy of Sciences
  \city{Beijing}
  \country{China}}
}
\affiliation{%
  \institution{
  University of Chinese Academy of Sciences
  \city{Beijing}
  \country{China}}
}

% \affiliation{ 
%   \position{$^1$Laboratory for Internet Software Technologies, $^2$State Key Laboratory of Computer Sciences,}
%   \department{Institute of Software Chinese Academy of Sciences, Beijing, China\\
%   $^3$University of Chinese Academy of Sciences, Beijing, China
%   }
% }

% \affiliation{%
%   \institution{Institute for Clarity in Documentation}
%   \streetaddress{P.O. Box 1212}
%   \city{Dublin}
%   \state{Ohio}
%   \country{USA}
%   \postcode{43017-6221}
% }

% \author{Zhiyuan Chang$^{1,3,8}$, Mingyang Li$^{1,3,8}$, Junjie Wang$^{1,3,*}$, Qing Wang$^{1,2,3,*}, Shoubin Li$^{1,2}$}

% \affiliation{ 
%   \position{$^1$Laboratory for Internet Software Technologies, $^2$State Key Laboratory of Computer Sciences, }
%   \department{Institute of Software Chinese Academy of Sciences, Beijing, China; \\
%   $^3$University of Chinese Academy of Sciences, Beijing, China; $^*$Corresponding author\\
%   $^8$Both authors contributed equally to this research.\
%   }
% }

\begin{abstract}
% \jie{comment}
%我觉得第一句话和后面衔接的不太好，而且第一句信息量也不大，现在这个题目似乎已经不需要了。没有系统测试什么事了。
%这里可以看看要不要再加一句话，我先写一句，后面可以再改
Natural language (NL) documentation is the bridge between software managers and testers, and NL test cases are prevalent in system-level testing and other quality assurance activities. 
Due to reasons such as requirements redundancy, parallel testing, and tester turnover within long evolving history,  there are inevitably lots of redundant test cases, which significantly increase the cost. 
Previous redundancy detection approaches typically treat the textual descriptions as a whole to compare their similarity and suffer from low precision. 
Our observation reveals that a test case can have explicit test-oriented entities, such as tested function Components, Constraints, etc; and there are also specific relations between these entities. This inspires us with a potential opportunity for accurate redundancy detection. 
% \yuan{comment}
%王老师更改部分
% Due to reasons such as requirements redundancy and parallel testing, there are redundant test cases, which significantly increase the testing cost and maintenance effort as software evolves.
% Previous redundancy detection approaches are typically based on the similarity comparison of the whole textual descriptions and suffer from low precision.
% Our observation on industrial NL test cases reveal the potential opportunities for accurate redundancy detection, i.e., test cases having explicit test-oriented entities, and the practical challenges in doing so, i.e., the complex relations associated with these entities.
% To tackle this, we propose a fine-grained approach {\tool} which extracts and compares the test-oriented entities \jie{under their involving relations}, for redundancy detection.
In this paper, we first define five test-oriented entity categories and four associated relation categories and re-formulate the NL test case redundancy detection problem as the comparison of detailed testing content guided by the test-oriented entities and relations. 
Following that, we propose {\tool}, a fine-grained approach for redundant NL test case detection by dissecting test cases into atomic test tuple(s) with the entities restricted by associated relations.
To serve as the test case dissection, {\tool}  designs a context-aware model for the automatic entity and relation extraction.
Evaluation on 3,467 test cases from ten projects shows {\tool} could achieve 91.8\% precision, 74.8\% recall, and 82.4\% F1, significantly outperforming state-of-the-art approaches and commonly-used classifiers.
This new formulation of the NL test case redundant detection problem can motivate the follow-up studies to further improve this task and other related tasks involving NL descriptions.

\end{abstract}

\ccsdesc[500]{Software and its engineering~Software testing and debugging}
\ccsdesc[500]{Software and its engineering~Acceptance testing}

\keywords{Test Case Redundancy, Entity and Relation Extraction, Natural Language Processing}

\maketitle

\section{introduction}
\label{sec:introduction}

Software testing is an activity to ensure that an entire system meets its requirements \cite{briand2002uml}.
% \yuan{comment}
% 王老师两次修订中都提到这个里不用强调System-level，这里需要删掉这句话吗
In the testing phase, testers need to analyze the requirements specification, identify all the test execution scenarios, and then instantiate them in manually written test cases \cite{wang2020automatic}.
Such test cases are typically described in natural language (NL).
Due to their adjustability and interpretability, the NL test cases are still prevalent in industrial practice \cite{li2020clustering}.
% In the industrial practice, acceptance test cases are usually specified by different employees.
% Due to redundancy of requirements, parallel testing, or distributed testing, a test suite may contains redundant test cases \cite{engstrom2013test}. 

A requirement covers multiple features, and there may be overlapping features among requirements.
For a large software project, the requirements are typically tested by different engineers, and engineers are not aware of the feature overlapping.
Test redundancy may produce when each test engineer individually designs test case(s) for assigned requirements \cite{engstrom2013test,marijan2018practical}.
% Because of the redundancy of requirements, parallel testing and other reasons, there may produce redundant test cases in a test suite \cite{engstrom2013test,marijan2018practical}. 
As the system evolves, the redundant test cases significantly increase the cost of testing, as well as maintenance effort\cite{marijan2018practical}.
The problem is especially obvious in the manual testing scenario where human testers must read through test steps and carry them out manually by interacting with the system \cite{hsu2009mints}.

To alleviate the issue, information retrieval-based approaches have been proposed to automatically detect redundancy among the NL test cases \cite{tahvili2019automated,li2020clustering,viggiato2021identifying}.
The general idea is to vectorize the description of the test case with text representing models, e.g., vector space model or Doc2Vec, and conduct the similarity comparison on it.
% After that, test cases with similar descriptions will be considered as redundant.
However, these existing approaches suffer from low accuracy because they treat test cases' textual descriptions as a whole, and thus can not capture its fine-grained semantic information and inherent meaning. 
% \rev{of their weakness in capturing test case's complete semantic information and inherent meaning. }
Meanwhile, we have the following two observations which can facilitate the similarity comparison and redundancy detection of the NL test case.
% \rev{the similarity-based approaches suffer from low precision due to the following two reasons.}

% \jie{comment}
%我这块仔细想了想，我觉得Similar test cases are not always redundant这个事情是个非常common sense的事情，不用这么细说，我们在上面一段大概提一下就行了。
%我觉得我们上面应该是说单纯的文本匹配会带来很多噪声，导致现有的研究效果不好。然后我们发现了test case里面有一些属性能够让这个redundancy这个工作做得更好。
%第一个就是 我们发现test cases里面有着比较明显的语义单元（或者是测试相关的内容单元，或者其他说法，反正就是entity这种），能够更好的区分测试用例是否是redundant，然后就类似于下面的例子，说有测试对象是啥，测试操作是啥，我们通过分别的对比，会发现他们是两个不同的测试用例，而如果只是单纯的文本匹配效果会不好。
%我觉得这样写还有个好处，就是能潜在的让读者知道，为啥要定义那些类型，否则的话可能读者get不到为啥要定义这些实体和关系。这里的分析相当于给出了一些例子，让读者能够更好的读懂为啥要做entity extraction，以及为啥分了这些类型。
% \textit{Similar test cases are not always redundant.}

\begin{figure}[htbp]
\centering
\includegraphics[width=0.95\columnwidth]{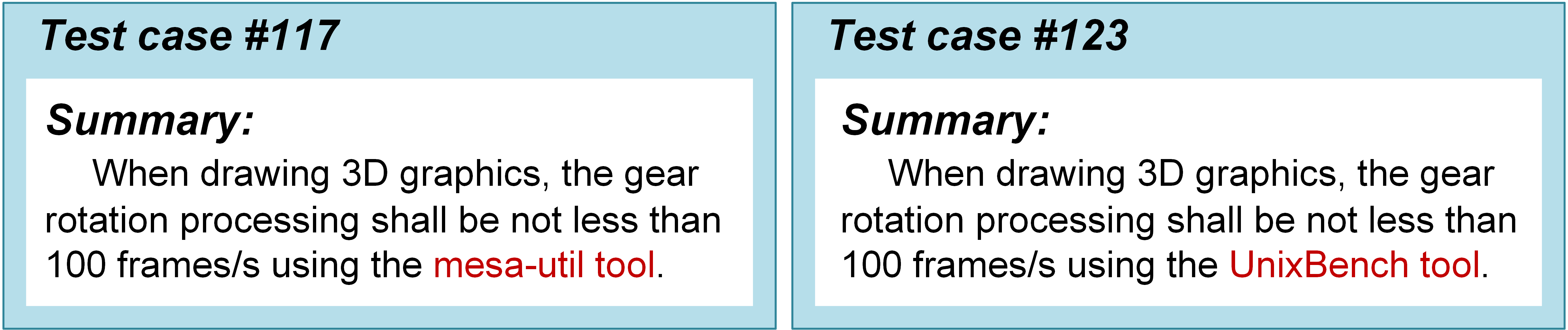}
\caption{Non-redundant test cases with similar descriptions}
\label{fig:similar_tcs}
\end{figure}

\textbf{First, the test case has explicit categories of test-oriented entities which can facilitate accurate redundancy detection.}
Take Figure \ref{fig:similar_tcs} as an example, the two test cases look similar in their textual descriptions, and would be detected as redundancy with the aforementioned information retrieval-based approaches.
However, if putting these two test cases under the microscope, we can find that the executing manners of these two test cases (``mesa-util tool'' and ``UnixBench tool'') are different, based on which, we can distinguish them accurately. 
% \yuan{comment} \jie{the executing manners of these two tests? comment}
%是不是用tests更好，就是这个测试的执行方式？不是这个测试用例的。或者你觉得怎么改
%这里还没有介绍到manner这个类别，如果没有举例“mesa-util tool” 和“unixbench tool”，会不会对这个executing manner的理解有歧义，我第一眼看以为是用不同的方式执行测试用例，而不是测试用例里面的测试工具描述
More than that, one can easily observe that there are different categories of test-oriented entities, for example, ``gear rotation processing'' is the tested functional component, while ``when drawing 3D graphics'' is the pre-conditions for executing the test case. 
Only when the specific categories of test-oriented entities are mapped, can the two test cases be determined as redundant. 
Taken in this sense, this paper aims at identifying the test-oriented entities to facilitate the accurate detection of NL test cases.

% \jie{comment}
%我重写了下，我感觉原来那个太啰嗦了，尤其是前半部分说已有研究不行，我感觉没必要那么大的篇幅，感觉那块非常的common sense

% ### to illustrate the necessity of entity extraction####
% For the previous studies, the basic assumption is that all the words in a test case play an equal role when vectorizing (full-text similarity).
% Under the assumption, if most of the words in two test cases are overlapping, they will be considered as redundancy.
% In practice, we observe that the assumption fails since some non-redundant test cases are distinguished by the test-oriented entities rather than the whole descriptions.
% For example, Figure \ref{fig:similar_tcs} shows the summaries of two non-redundant test cases in the operating system testing.
% Since they are both to test the performance of ``gear rotation processing'', the descriptions are similar.
% The difference is the testing manner (\#117 uses ``mesa-util tool'' while \#123 uses ``UnixBench tool''), and there are only a few words to express this difference in the descriptions.
% Nevertheless, 
% Taken in this sense, this paper aims at identifying the test-oriented entities, e.g., the tested functional component and the behavior operating on it, for the accurate redundancy detection.
% \jie{comment}
%这段可能还得再调整下，我觉得基本思路应该是说，前面一段已经说明白了已有研究不work，感觉reviewer也能一目了然的纸袋为啥已有研究不行，所以我感觉这个层面不应该细说。我觉得这里应该细说的是，我们从中发现了哪些现象，这些现象能够帮助我们进行redundancy detection

% \jie{comment}
%承接着上面那一条的修改，我觉得这里我们是不是要说，除了基本的测试内容单元，还会存在一些关系，也需要进行区分。或者说一个测试用例会存在多个测试内容单元，以及附带的关系，需要进行精确的识别和区分。
%因为我感觉现在multiple这个事情，好像就是coverage rule那一小部分对应这个事情？在redundancy detection中占大部分的XX comparison并不太是对应这个multiple feature的，这些XX comparison只是为了让匹配的更准确而已？
%所以我觉得这条还不如从关系的角度来说？正好能对应到那里的实体和关系的定义。后面的redundancy detection就是为了匹配更好。
%第二条这里我不太清楚这样是不是比较妥当，你们再看看。但是第一条我觉得应该这么整。
%而且其实这个multiple，也就是说要特别准确的匹配，其他的好像也没有带来方法上面的其他考虑。精确匹配了，这个就能识别好了。
% \textit{There may be multiple tested features within a test case.}

\begin{figure}[htbp]
\centering
\includegraphics[width=\columnwidth]{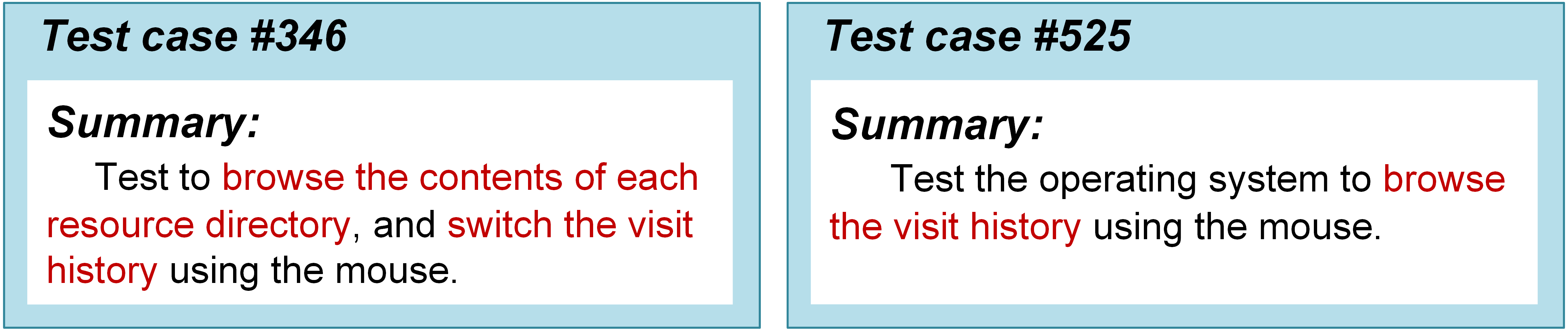}
\caption{Non-redundant test cases with multiple test-oriented entities}
\label{fig:granularity}
\end{figure}

\textbf{Second, there might be multiple test-oriented entities that need to be carefully parsed and matched to ensure accurate redundancy detection.}
% \jie{Can we add an example to illustrate the necessity to use the relation? For this example, it only implies that we should accurate conduct the matching.}
% \textit{Test case is not an atomicity element and redundancy is directional due to different test granularity.}
% due to different test granularity .}
% ### to illustrate the necessity of relationship extraction ###
The first observation has motivated us to conduct the comparison within the same category of test-oriented entities for determining redundancy. 
However, when we put the two test cases in Figure \ref{fig:granularity} under the microscope, a second observation is made.
There are both testing \textit{Behavior} ``browse'' and tested \textit{Component} ``visit history'' in these two test cases, yet they are expressing different test-oriented operational information. 
In detail, in test case \#346, the \textit{Behavior} ``browse'' is targeted at \textit{Component} ``content of each resource diretory'', and the \textit{Component} ``visit history'' is associated with the \textit{Behavior} ``switch'', while in test case \#525  \textit{Behavior} ``browse'' is directly for \textit{Component} ``visit history''.
The observation implies that the multiple test-oriented entities need to be carefully parsed and matched, and it is necessary to identify the test-oriented operational information, i.e., entities and associated relations when analyzing test cases to achieve accurate redundancy detection.

Motivated by the two findings, we define five test-oriented entity categories, i.e., \textit{Component}, \textit{Behavior}, \textit{Prerequisite}, \textit{Manner} and \textit{Constraint}, and four relation categories associated with the entities.
We then re-formulate the NL test case redundancy detection problem as the comparison of detailed testing content guided by the test-oriented entities and relations. 
% the dissection of the test case into atomic test unit(s) with the five entities guided by their associated relations, and similarity comparison on them. 

% \jie{how about `Tscope'? or can upper case more character. it might be more informative.} 
 %\jie{we re-formulate the redundant test case detection problem and }  
Following that, we propose a fine-grained redundant test case detection approach {\tool}\footnote{We name our approach as {\tool} considering it likes a microscope to inspect the detailed information in test cases to facilitate the redundant detection.}, which dissects the test case into atomic test tuple(s) with the five entities restricted by their associated relations, and conducts the comparison on them. 
One example test tuple dissected from Test case \#525 in Figure \ref{fig:granularity} is as follows, \textit{Behavior} ``browse'', \textit{component} ``visit history'' and \textit{Manner} ``mouse''. 
% automatically extract the test-oriented entities and their associated relations, dissect each test case into structured atomic testing units, and conduct the following  similarity comparison.
% \jie{Its key idea is to extract the test-oriented entities in terms of categories and their associated relations, and compare the similarity based on the fine-grained test-oriented information. }
% Its key idea is to extract the test-oriented entities and their associated relations, dissect each test case into structured atomic testing units, and conduct the similarity comparison based on it.
% \jie{comment}
%key idea这句话，我没有特别想好，decompose into tuple 这个事情要不要在这句话里面说，我感觉这是一个非常细节的事情，而这里感觉应该是从稍微high level说下基本思路是啥。
%我觉得不需要一直强调和已有研究不同，不是full text这些，我觉得你把你的思路说明白，人家一目了然的就知道你是怎么做的了。
% Specifically, we define five test-oriented entity categories, i.e., \textit{Component}, \textit{Behavior}, \textit{Prerequisite}, \textit{Manner} and \textit{Constraint}, and four relation categories associated with the entities.
To achieve this, {\tool} first designs a context-aware model for extracting test-oriented entities and relations from test case descriptions, which considers the global context of the test case for entity extraction, and the local context of the involving entities for relation extraction. 
After that, {\tool} dissects each test case into the structured atomic testing tuple(s) guided by the extracted entities and relations. 
% for better conducting the fine-grained comparison.
% Each structured {\tuple} consists a \textit{Component}, an associated \textit{Behavior}, an associated \textit{Manner}, an associate \textit{Pre}, and an associated \textit{Manner} for fine-grained comparison. 
% \jie{i.e., the structured test-oriented operation}. \jie{comment}
%structured or explainable？我就是觉得这里只说tuple，让人一眼不知道，是做了个啥，为啥要这样做呢。所以感觉这里还是得让人一眼就知道原理
%或者这里加个example？或者是在前面那段（第二个例子）就大概提一下这个意思
Finally, {\tool} detects redundancy by comparing the entities in each tuple pair,  considering the semantic meaning of the entities as well as their involved indicative words.

We evaluate {\tool} on 3,467 test cases from ten projects.
% \jie{comment}
%这里要不要弱化一下，不说domain了。只有一个domain感觉会被质疑。在evaluation那块也可以先不说了。
The evaluation results show that {\tool} could reach 97.5\% precision, 94.8\% recall for the entity extraction, and 90.4\% precision, 97.6\% recall for the relation extraction, which significantly outperforms two state-of-the-art approaches. 
For the redundancy detection task, {\tool} could achieve 91.8\% precision, 74.8\% recall and 82.4\% F1.
Compared with the two state-of-the-art redundancy detection approaches and four commonly-used classifiers, {\tool} is 19.8\%-23.4\% higher in F1.
Moreover, the results of ablation experiments show that the five entity categories all play significant roles in {\tool}.

The new formulation of the NL test case redundant detection problem can motivate the follow-up studies to further improve this task, and other related tasks involving NL descriptions. 
Actually, there are several tasks in software engineering domain involving the similarity comparison of two textual documents, e.g., duplicate test reports detection \cite{isotani2021duplicate,huang2020quest}, similar Stack Overflow questions identification \cite{wang2019detecting}, duplicate requirements detection \cite{motger2020resim}, etc. 
The previous techniques typically treat the textual descriptions as a whole for the similarity comparison, while ignoring the fine-grained semantic information hidden in the text. 
% \yuan{comment}\jie{comment} 
%我觉得inherent structured information挺好的，王老师好像觉得不行。那可以改成fine-grained semantic information
%这里提到inherent structured information 和上面的描述不太一致，这个说法需要和上面说fine-grained semantic information统一吗
The new formulation proposed in this paper, i.e., comparison of detailed content guided by the scenario-related entities and relations, could potentially motivate the researchers in these related fields.

In summary, the key contributions of this paper are as follows:
\begin{itemize}
    \item
    The new formulation of the NL test case redundancy detection problem, i.e., the comparison of detailed testing content guided by the test-oriented entities and relations.
    % the dissection of test case into atomic test unit(s) with five entities guided by their associated relations, and similarity comparison on them. 
    \item 
    A fine-grained redundancy detection approach {\tool} for NL test cases, which dissects the test case into atomic test tuple(s) with the five entities restricted by their associated relations, and conducts the comparison on them.
    % compares the test case based on their entities guided  by their relations. 
    \item
    A context-aware model for extracting test-oriented entities and their relations from test case descriptions, which involves the global context of the test case in entity extraction, and the local context of the involved entities for relation extraction.
    % \jie{comment}
    %我觉得没必要说elaborating a state-of-the-art information extraction model in the NLP field，就是说你的特点是啥就行。
    %includes a novel BERT+Attr-CRF model to automatically extract the fine-grained phrases (i.e., problematic features). It combines the review descriptions and review attributes (i.e., app category and review description sentiment) to better model the semantics of reviews and boost the performance of the traditional BERT-CRF model 这是之前文章里面写的，可以参考下。但可能还可以写得更简练和有趣一些吧
    \item
    Evaluation with 3,467 test cases from ten projects, with promising results. We also publicize the source code\footnote{https://github.com/czycurefun/testcase$\_$detection} for facilitating follow-up studies and other related tasks. 
\end{itemize}
% \jie{comment}
%第一条和第二条也可以考虑互换位置。我全部改成了短语，不是主语+谓语，这样比较简洁。
%可能少量和前面不一致的地方，主要就是分解成tuple相关的，可以再微调下。

% \jie{example papers are in the source file, and can add more.}
%Haruna Isotani, Hironori Washizaki, Yoshiaki Fukazawa, Tsutomu Nomoto, Saori Ouji, Shinobu Saito: Duplicate Bug Report Detection by Using Sentence Embedding and Fine-tuning. ICSME 2021: 535-544
% Yuekai Huang, Junjie Wang, Song Wang, Zhe Liu, Yuanzhe Hu, Qing Wang: Quest for the Golden Approach: An Experimental Evaluation of Duplicate Crowdtesting Reports Detection. ESEM 2020: 17:1-17:12

% Liting Wang, Li Zhang, Jing Jiang: Detecting Duplicate Questions in Stack Overflow via Deep Learning Approaches. APSEC 2019: 506-513

% 	Quim Motger, Cristina Palomares, Jordi Marco: RESim - Automated Detection of Duplicated Requirements in Software Engineering Projects. REFSQ Workshops 2020

The remainders of the paper are as follows:
Section \ref{sec:background} presents the empirical studies of the entity category for redundancy detection.
Section \ref{sec:approach} elaborates the approach. 
Section \ref{sec:experiment} presents the experiment design. 
Section \ref{sec:result} describes the results.
Section \ref{sec:discussion} discusses the learned lessons. 
Section \ref{sec:related work} introduces the related work and its limitations. 
Section \ref{sec:conclusion} concludes our work.

\section{Empirical analysis of Entities and Relations}
\label{sec:background}
% This section presents the definitions of the test-oriented entity categories and their relation categories, and the correlation analysis results between the entities and redundancy detection.
% \jie{comment}
%这部分能否分成两个小章节，一个叫做taxonomy of entity and relation in test case，另一个叫做 我还没想好 ，我稍微觉得这块有点很common sense，如果两个被测对象相同，那肯定他们就有很大概率相同啊，我想想如果组织这块更好
%前面entity and relation 介绍那块，要不要给个具体的例子，我感觉只看这个表格，我都不太知道这个means，standards是个啥意思。尤其是entity这块。而且这个means叫法合适吗，能不能叫manner？当然这里我不确定，我只是觉得means我以为是意味着这种；然后standards特别像是标准，能否叫做constraint

\subsection{Categories of Entities and Relations}

\begin{table*}[htbp]

  \caption{The entity and relation categories}

  \label{tab:entity_category}
\resizebox{\textwidth}{!}{
  \begin{tabular}{lll|l}
    \toprule
~  &  \textbf{Category} & \textbf{Definition}   &   \textbf{Examples} \\   
      \midrule
 ~  &   Component (Com)  &  the tested functional component & \multirow{5}{*}{\begin{minipage}[b]{\columnwidth}
		\centering
		\includegraphics[width=\linewidth]{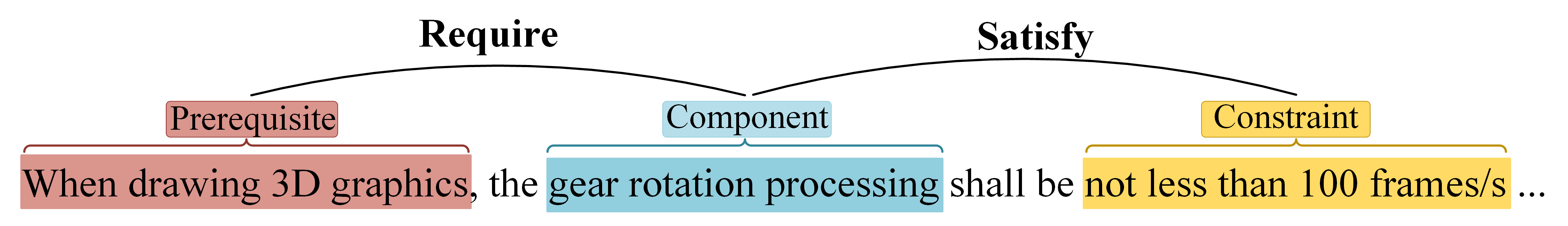}
	\end{minipage}}    \\ 
 ~  &   Behavior (Beh) &  the behavior acting on the tested component &  \\ 
\textit{Entity} &  Prerequisite (Pre) & the pre-conditions before the test case is executed  & \\
 ~  &   Manner (Man) &  the executing manner of the test case & \\
 ~  &   Constraint (Con) &  the constraints to be met after the test case is executed  &  \\

\cmidrule(lr){1-3}
% \midrule

 ~  &   Act (Act)  &   the relation between \textit{Component} and \textit{Behavior}  & \multirow{4}{*}{\begin{minipage}[b]{0.85\columnwidth}
		\centering
		\includegraphics[width=\linewidth]{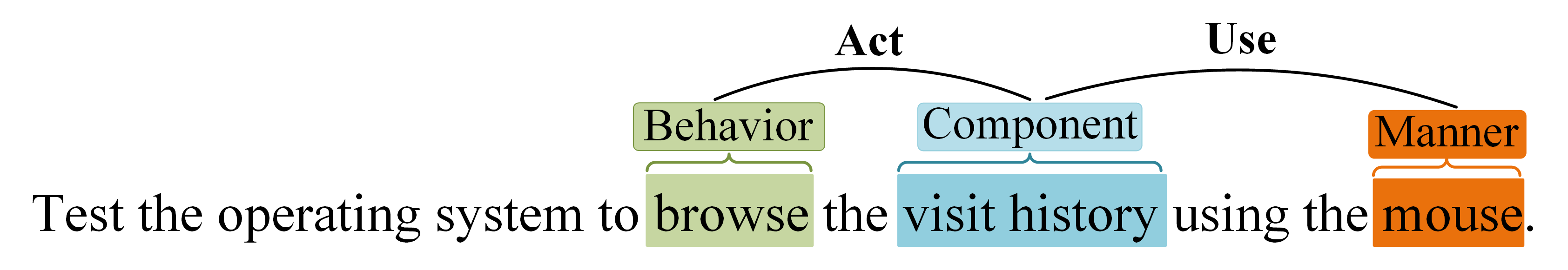}
	\end{minipage}}  \\
\textit{Relation}   &    Require (Req) &  the relation between \textit{Component} and \textit{Prerequisite}    &   \\ 

 ~  &   Use (Use) &   the relation between \textit{Component} and \textit{Manner}    &  \\
 ~  &   Satisfy (Sat) &  the relation between \textit{Component} and \textit{Constraint}    &   \\

\bottomrule
\end{tabular}
}
\end{table*}

Motivated by the observations in Section \ref{sec:introduction}, we provide a new formulation of the NL test case redundancy detection problem, i.e., the comparison of detailed testing content guided by the test-oriented entities and relations.
% the similarity comparison of the test-oriented entities with their associated relations. 
To achieve this, we define five categories of entities and four categories of relations associated with the entities.
Specifically, we explore entity and relation categories through a bottom-up analysis approach.
Specifically, three researchers (details in Section \ref{experiment:dataset}) are involved in mining the categories of entities and relations that affect redundancy detection in the test case text.
If all three researchers agree on adding a category, this entity category is admitted and added to the entity category set.
While if their views diverge, the decision is made through a voting mechanism, i.e., the entity category will be added to the set if it is admitted by at least two researchers.
Finally, we obtain the five entity categories and corresponding relations among the entity categories.
Table \ref{tab:entity_category} shows each entity/relation category and examples.

\subsubsection{\textbf{Categories of Entities}}
The definition of the five entity categories is based on the purpose and basics of software testing, as well as the observations on NL test cases.
First, test cases are driven by the feature(s) in requirements, and a feature specifies the \textbf{behavior} of one or more \textbf{components} in terms of their current \textbf{conditions} \cite{cartaxo2007test}.
Taken in this sense, the key entities in a feature will also be reflected in the test case descriptions.
Therefore, we identify three entity categories ``Component'', ``Behavior'' and ``Prerequisite'' respectively.

Second, according to our observations, test cases differ by the \textit{Manner} sometimes.
For example, there are descriptions of two non-redundant test cases in Figure \ref{fig:similar_tcs}.
The two test cases have the same \textit{Prerequisite} (``When drawing 3D graphics'') and \textit{Component} (``gear rotation processing''), but different operation manner (``mesa-util tool'' and ``UnixBench tool'').
To reflect this difference, we define an entity category ``Manner''.
% \jie{comment}
%我觉得可以把constraint去掉，因为这里还没介绍到constraint呢。

Third, in some cases, test cases may differ by the satisfied constraints.
For example, there are two descriptions, ``Test there are preset applications after the system installation'' and ``Test the preset applications including FTP application after the system installation''.
The two test cases have the same \textit{Component (``preset applications'')} but the latter additionally involves the constraint (``including FTP application'').
Accordingly, we define an entity category ``\textit{Constraint}'' to indicate the difference.
% \jie{comment}
%这个constraint能用表格里面的吗，就是not less than 100 这个，现在这个例子都没有出现过。
%这里因为图1中的两个例子是都有contraint的，所以不能说明constraint的重要性。

\subsubsection{\textbf{Categories of Relations}}
As shown in Figure \ref{fig:granularity}, there may be multiple test-oriented entities per entity category within a test case, which implies the need for inspecting the entities within the test case a step further.
Taking Test Case \#346 in Figure \ref{fig:granularity} as an example, \textit{Behavior} ``browse'' is targeting at \textit{Components} ``contents of each resource diretory'', and \textit{Behavior} ``switch'' is acting on \textit{Components} ``visit history''.
This demonstrates the mapping between \textit{Components} and \textit{Behavior}, and we define it as the \textit{Act} relation.

We also observe the relations in terms of the other three categories of entities, e.g., the executing manner of the testing. 
And considering the components in the test case are the basic object of the testing content, we define other three relations between \textit{Component} and \textit{Prerequisite}, \textit{Manner}, \textit{Constraint} to indicate the detailed information of the testing (details in Table \ref{tab:entity_category}).

% \jie{comments}
%这块哪些需要倾斜，哪些需要加引号，哪些需要大写这种，志远再稍微调整下哈。
%我觉得在table1中，就直接用动词原形就行吧，不需要加s吧
%已修改和添加
% there are four \textit{Components} (point, line, rectangle and image color), two \textit{Behaviors} (draw and adjust) and a \textit{manner} (drawing editor).
% It 
% we also define four relation categories to determine association between the \textit{Component} and the other four entity categories.
% Table \ref{tab:entity_category} shows the four relation categories.

% \jie{comment}
%例子不对。
%这里我有个疑问，为啥要以component为中心的，为啥抽取component相关的呢，acts肯定是的，感觉这个prerequite是针对这个测试内容的，不算是针对component的，manner，constraint也是。感觉只有acts是一个里面存在多个对应，是需要抽取的。但我似乎觉得其他三种是针对一个component-behavior对的。

\subsection{Correlation Analysis}
\label{sec:correlation_analysis}
We conduct an empirical study to investigate the effectiveness of the entity categories for redundancy detection.
Specifically, we randomly sample 5,000 test case pairs and manually label each test case by comparing each pair.\footnote{The test case pairs are built from the dataset in Table \ref{tab:dataset}. The pairing and labeling processes are consistent with the descriptions in Section 4.2.}
Then, we build five Boolean variables by manual judgment, i.e., $EQ_{com}$, $EQ_{beh}$, $EQ_{pre}$, $EQ_{man}$ and $EQ_{con}$.
Each variable represents the entities belonging to each category in the summaries are manually judged as equivalent.
% \begin{itemize}
%     \item 
%     $EQ_{com}$: whether the entities belonging to `Component'' in the test case pair are equivalent;
%     \item
%     $EQ_{beh}$: whether the entities belonging to `Behavior'' in the test case pair are equivalent;
%     \item
%     $EQ_{stat}$: whether the entities belonging to `State'' in the test case pair are equivalent;
%     \item
%     $EQ_{mea}$: whether the entities belonging to `Manner'' in the test case pair are equivalent;
%     \item
%     $EQ_{stan}$: whether the entities belonging to `Constraint'' in the test case pair are equivalent.
% \end{itemize}
At the same time, a variable $Redundant$ is built according to the redundancy label (not based on entity comparison), representing whether a test case is truly redundant.
% The variable $Redundant$ represents 
% We calculate the Cohen\' kappa coefficient and analyze the consistency of the distribution of two variables, $Equ_{entity}$ and $Redundant$.

\begin{table}[htbp]
  \caption{The correlation for each entity category}
  \label{tab:corelation}
\resizebox{0.9\columnwidth}{!}{
\begin{tabular}{cccccc}
\toprule
\textbf{Variable} & \textbf{ $EQ_{com}$} &   \textbf{ $EQ_{beh}$}    &   \textbf{ $EQ_{pre}$}   & \textbf{ $EQ_{man}$}  &   \textbf{ $EQ_{con}$}\\   
\midrule

$Redundant$ &   0.509   &   0.695   &   0.841   &   0.632   &   0.707\\
 $p-value$ &   <0.001   &   <0.001   &   <0.001   &   <0.001   &   <0.001\\

\bottomrule
\end{tabular}
}
\end{table}
% information grain \cite{kent1983information} of the
We analyze the correlation between the above five variables and the variable $Redundant$.
%The Pearson correlation coefficient between $Redundant$ and above five variable are 0.509, 0.695, 0.841, 0.632 and 0.707 respectively, and p-values of the correlation test are all less than 0.05.
Table \ref{tab:corelation} shows the Pearson correlation coefficient and p-value of the correlation test.
The results show that the five entity categories are significantly correlated to the variable $Redundant$, which indicates the effectiveness of each entity category for redundancy detection.
Moreover, we analyze the consistency of the two variables, i.e., $EQ_{all}$ and $Redundant$, where $EQ_{all}$ represents that the entities belonging to the five entity categories in the test case pair are all equivalent by manual comparison.
Cohen\' kappa coefficient is 0.984, which shows the significant consistency of the two distributions.
The results indicate that redundant test cases could be effectively detected using the five entity categories.
Motivated by the above considerations, we design a joint extraction model to extract entities and relations belonging to the pre-defined categories, dissect each test case into atomic test tuple(s) based on the extracted entities and relations, and detect the redundancy based on them, rather than considering the textual descriptions as a whole like previous approaches.

\section{approach}
\label{sec:approach}
\begin{figure*}[htbp]
  \includegraphics[height=5cm,width=\textwidth]{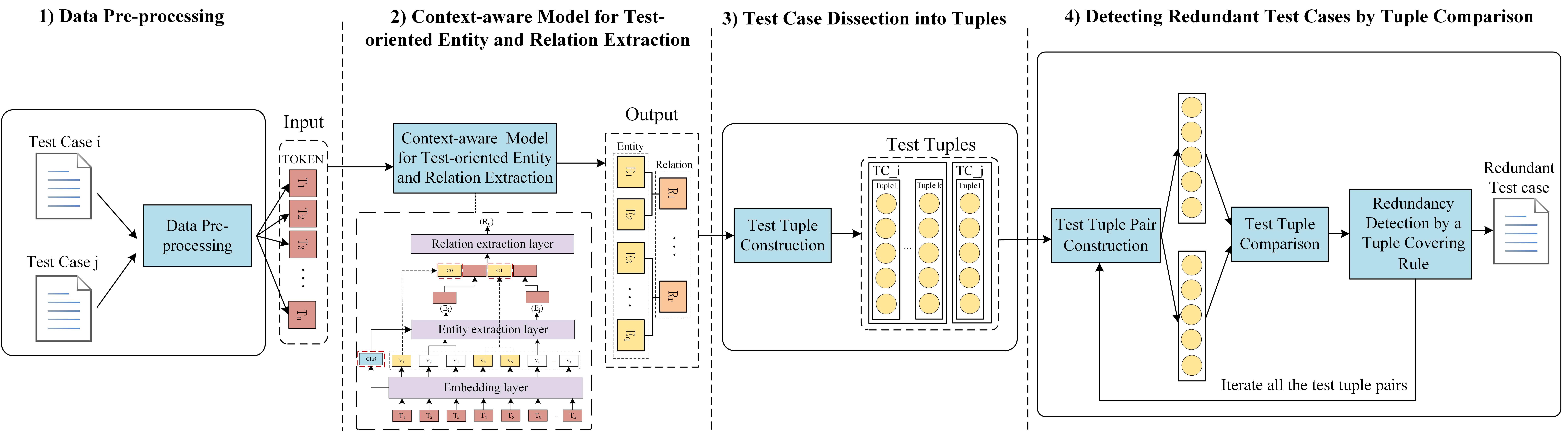}
    \caption{
     The overview of {\tool}
    }
    \label{fig:approach_overview}    
\end{figure*}

Figure \ref{fig:approach_overview} shows the overview of {\tool}.
{\tool} consists of four phases: (1) Data Pre-processing, where it conducts data-processing and constructs samples for the extraction model; (2) Context-aware Model for Test-oriented Entity and Relation Extraction, where {\tool} designs a context-aware extraction model to extract the test-oriented entities and relations from test case descriptions; (3) Test Case Dissection into Tuples, where {\tool} dissects each test case into test tuples based on the extracted entities and relations, to represent the fine-grained test-oriented operational information;
%第三步我还是建议更清晰的写出来拆分是干啥的，拆分出来的tuple含义是啥。
% decompose each test case into tuple(s) based on the extracted entities and relations, to represent the atomic testing operational information , 或者 atomic testing content ，或者再想想。
%我看到现在intro里面没有说这个tuples，用的是structured atomic testing units，我稍微有点感觉这个atomic 和 units有些重复了，要不就简单点 structured atomic testing content？
%要不我就觉得会显得很细节，就是做了拆分，不知道这个rational是啥
%而且我们题目里面有个fine-grained，那这里的描述里面要不要稍微提一下这个fine-grained，我们认为的fine-grained，是说抽取实体和关系是fine-grained，还是这个拆解元组是fine-grained，还是总体上是。感觉这个关键词可以出现一下。
and (4) Detecting Redundant Test Cases by Tuple Comparison, where {\tool} designs three comparison strategies for tuple comparison and detects redundancy by a \textit{Tuple Covering Rule}.
The following introduces the details of the four phases.

\subsection{Data Pre-processing}
\label{sec:data_preprocessing}

Considering test cases are written in natural language, {\tool} applies the standard data pre-processing pipeline in NLP field \cite{nayak2016survey}.
Specifically, given a test case $TC$, {\tool} first splits the textual contents into sentences $[s_1, ..., s_i, ..., s_n]$.
For a sentence $s_i$, {\tool} removes special characters, converses into lowercase, and tokenizes it into a token sequence using the NLP toolkit scikit-learn\footnote{https://scikit-learn.org/stable/}.
Then, each test case is represented as a token sequence $T_{TC}$ = [$T_1$, $SEP$, $T_2$, $SEP$, ..., $T_i$, $SEP$, ..., $T_n$], where $T_i$ is the token sequence for sentence $s_i$, \textit{SEP} is the placeholder for dividing sentences.
After that, $T_{TC}$ is considered a sample for entity and relation extraction.

\subsection{Context-aware Model for Test-oriented Entity and Relation Extraction}
\label{approach:extraction_model}

We adapt the entity and relation joint extraction techniques \cite{zheng2016joint,dixit2019span,tan2020boundary} to design our context-aware model for test-oriented entity and relation extraction.
First, the model obtains the candidate entities by iterating all the spans \cite{tan2020boundary} in the input and encodes each candidate span using an embedding layer.
Second, it designs an entity classifier, which considers the global context of the test case, to determine whether each candidate is an entity and its category.
Third, it designs a relation classifier to decide the relation category for each entity pair, where it introduces the local context information of involved entities to act as the indicators for relation classification.

% \jie{comment}
%不过我写完之后，我觉得这可能也是个人习惯问题，就像mingyang先写通用做法，再写我们的新内容也行。
%但我还是建议1，分成span和encode合并，因为这两块不重要。而且我看下面的小章节，也就只有embedding
%后面我们的共性，就是global semantic和context information，最好再强调一下，分开写，写明白每个部分的作用。

% We design a joint extraction model following the model architecture in \textit{SLM} \cite{dixit2019span} which is the state-of-the-art joint extraction model in NLP field.
% SLM starts with an embedding layer, which encodes the each token in the input sample into a hidden representation. 
% Then, the hidden representations pass through an entity extraction layer and a relation extraction in sequence.
% Finally, the model outputs the extracted entities and their relations.
% In our stu

% Figure \ref{fig:extraction_model} shows the architecture of the joint extraction model.
% The joint extraction model starts with an embedding layer, which encodes the each token in the input sample into a hidden representation. 
% Then, the hidden representations pass through an entity extraction layer and a relation extraction in sequence.
% Finally, the model outputs the extracted entities and their relations.
% % Compared with SLM, the joint extraction model additionally introduces the context information into the two layers respectively.
% Following introduces the details of each layer.

\begin{figure}[htbp]
\setlength{\abovecaptionskip}{1pt}
\setlength{\belowcaptionskip}{1pt}
\begin{center}
  \includegraphics[height=7cm,width=9cm]{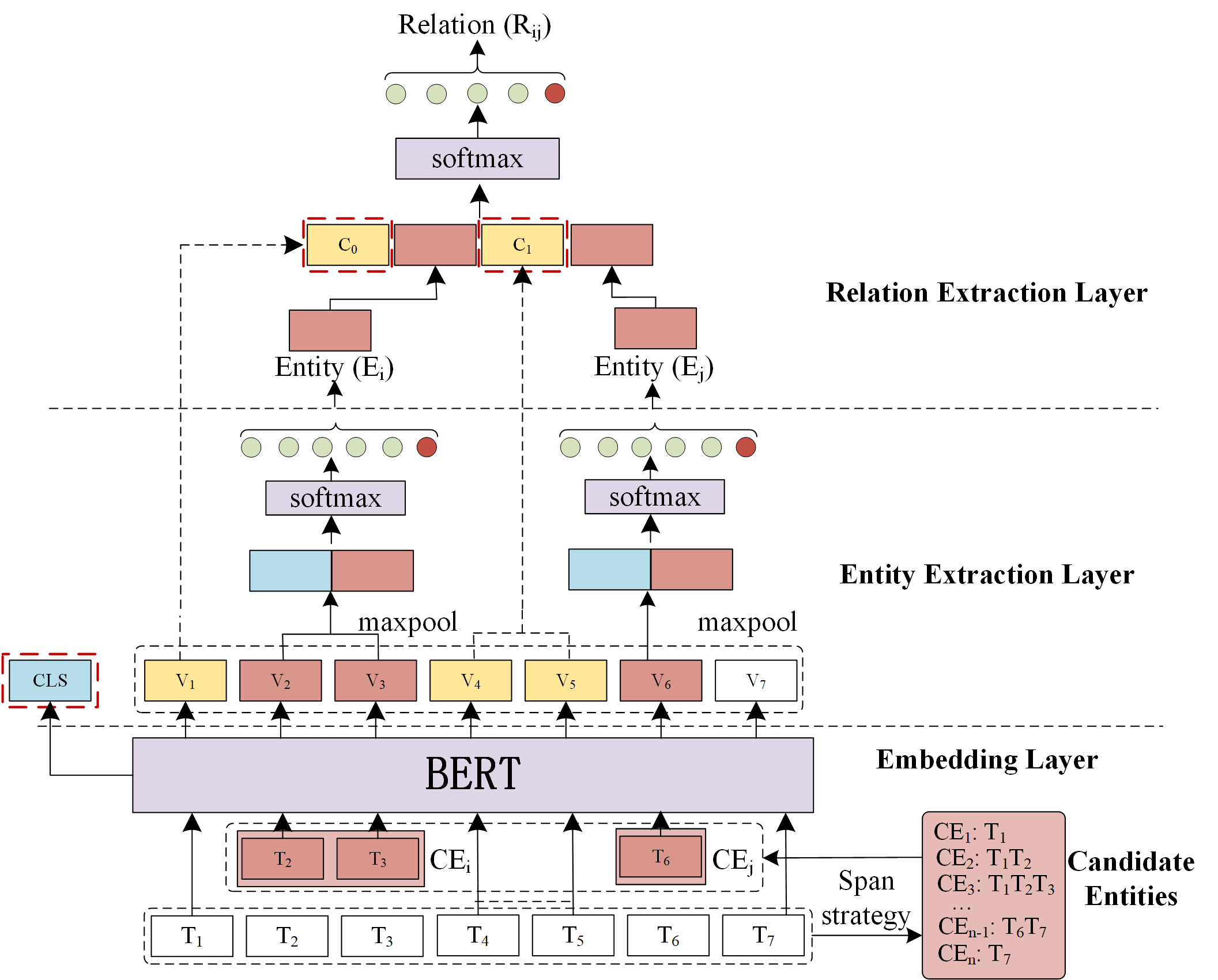}
  \end{center}
    \caption{
     The architecture of the context-aware model
    }
    \label{fig:extraction_model}    
\end{figure}

\subsubsection{\textbf{Embedding Layer}}
In this layer, the extraction model firstly iterates all the candidate spans \cite{tan2020boundary}.
Specifically, the extraction model presets a span length and constructs all the spans by traversing all consecutive word chunks in the input text that do not exceed the span length. 
Spans are regard as candidate entities ($CE_1$ to $CE_n$  in the Figure \ref{fig:extraction_model}).
Then, the embedding layer encodes each candidate entity into a hidden representation. 
Please note that the span length is empirically set as 10 in the extraction model since there are no more than ten words for most of the entities according to our observations. 
After that, our model uses a pre-trained BERT model\footnote{https://github.com/huggingface/transformers} which is a commonly-used embedding model in the NLP field and shows strong robustness for different domains \cite{abs-2003-11563,SunY19}.
% The reason for replacing BiLSTM with BERT model is that the BERT model has been pre-trained on a large volume of corpus, which could potentially reduce the number of required training samples.
% \jie{comment}
%我觉得前面那块，可以比较笼统的说我们用了个什么样的framework，可以引用几篇文章，不固定到某篇文章上面。然后这里就不用说我们replace BiLSTM with Bert，就说我们这层用的bert就行，因为这个也不算怎么大的贡献，也不是现在比较通用的做法了。
% In addition, the BERT model .
Through the embedding layer, {\tool} produces a hidden representation for each candidate entity.
% $V_{tc}$ = ($V_{t_1}$, $V_{t_2}$, ..., $V_{t_i}$,..., $V_{t_n}$) where $V_{t_i}$ is the embedding vector for each token in the token sequence $T_{tc}$ .

\subsubsection{\textbf{Entity Extraction Layer}}
% \jie{comment}
%我感觉这里说的和前面说的有点不太一致。感觉这里的详细描述 是说先做了embedding，然后组装了span。但在前面总体介绍里面，是先说了组织span 又说了embedding。我觉得需要保持一致。
%是不是可以把第一步叫做embedding and candidate entity  construction这种感觉的？到了entity extraction layer 就直说怎么抽实体？这块你们再看看，我不太确定。
%而且把candidate entity construction分在前一部分的话，如果是说先这里是不是可以叫做entity classification layer了？
%不过可能分开也不是很好，因为后面的relation那里，感觉candidate relation construction是放在了对应章节里面。所以candidate entity这个 可能也还是得方法entity extraction layer里面。那样的话，在前面的介绍部分就需要修改下。
%我又看了一下figure 4，确实这个candidate entity生成是在这个entity extraction layer里面的。
In this layer, {\tool} receives the hidden representations of all candidate entities and outputs whether they are entities and their categories.
For the hidden representation of $CE_i$, {\tool} leverages a max\-pooling function \cite{lakshmi2019human} to retain the key semantic information and filter the noise.
After that, it is sent to the entity classifier.
% \jie{i revise this para, and can be further improved.}

Compared with the traditional extraction models which only use the representations of the candidate entities for classification, the entity classifier in our model additionally includes the global context of the input test case to help determine the certain category of an entity belonging to.
% Specifically, for traditional models, the entity classifier only uses the local semantics $MP_{CE_i}$ for $CE_i$. 
% They pay more attention to the hidden semantic representation of the input text. 
The reason why we employ this global context is that different types of test cases would have unique nature in the test-oriented entities. 
For example, a test case targeting at the performance bugs would be more likely to have the \textit{Constraints} and \textit{Prerequisite} categories of entities, compared with the test case targeting at the scalability bugs.
% In practice, we observe that the global semantics the whole test case carries could benefit for entity classifier.
% For example, if a test case aims at testing the feature performance, it is more likely to describe \textit{Component}, \textit{Constraints} and \textit{Prerequisite}.
% While, for a test case tends to test the scalability of a component, \textit{Constraints}, \textit{Prerequisite} may appear less frequently. 
% For example, the \textit{Means} typically appears after the word ``using'' or ``by''.
% Not limited to the example, the context information usually provides clues for entity extraction.
Specifically, the model applies a vector $CLS$ \cite{qiao2019understanding} for signifying the global context. 
It is the weighted sum of hidden representations for all the tokens in the test case and has been proven to effectively improve the performance of the classification tasks\cite{xiong2021fusing}.
% \jie{comment}
%这块我也觉得不怎么好，感觉最后这句说 has been proven to indicate global semantic，感觉前面可以说global semantic，但这里has proven 引用其他文献了，能不能说得清楚一些，到底是啥
% % deep insights between test categories and entities in test cases. 
% Therefore, besides the $MP_{CE_i}$, the joint extraction model additionally introduces a vector $CLS$ \cite{qiao2019understanding} in the entity classifier component.
% $CLS$ is the weighted sum of $MP_{tc}$ for the whole test case, and has been proved to effectively indicate the semantic information that the whole text carries \cite{xiong2021fusing}.
% Based on the above considerations, the joint extraction model sends $MP_{CE_i}$ to an entity classifier together with $CLS$.
The entity classifier concatenates the representation of each candidate entity and $CLS$, and uses a softmax function \cite{Zunino2002Analog} to predict whether the candidate entity is an entity and its entity category.
The outputs of softmax function are six probabilities, i.e., $P_{Com}$, $P_{Beh}$, $P_{Pre}$, $P_{Man}$, $P_{Con}$ and $P_{Non}$, where the former five represent the probabilities that the candidate belonging to the entity categories, and $P_{Non}$ represents the probability that the candidate does not belong to any categories.
After that, the entity classifier chooses a category with the highest probability.
Finally, the entity extraction layer outputs all the entities belonging to the five entity categories.

\subsubsection{\textbf{Relation Extraction Layer}}
The relation extraction layer obtains the extracted entities from the entity extraction layer and judges the relation between each entity pair.
Take two extracted entities as $E_i$ and $E_j$, and the relation between the two entities as $R_{ij}$. 
Relation extraction layer also use a softmax function to predict the probabilities of five categories $R_{ij}$ belongs to, e.g., $P_{Act}$, $P_{Req}$, $P_{Use}$, $P_{Sat}$ and $P_{Non}$, where the former four represent the probability of the relation categories and $P_{Non}$ represents there is no relation between $E_i$ and $E_j$.

Traditional models only employ the representations of $E_i$ and $E_j$ as input.
However, we observe that besides the two entities, the local context information is beneficial for relation extraction.
Taking Test Case \#525 in the Figure \ref{fig:granularity} as an example, for the \textit{Component} ``visit history'' and \textit{Manner} ``mouse'', the context ``using'' could be a trigger word to indicate the \textit{Use} relation between the two entities.
Thus, when classifying the relation between $E_i$ and $E_j$, our model additionally introduces the local context information $C_0$ and $C_1$, where $C_0$ is the contextual words before $E_i$, and $C_1$ is the contextual words between $E_i$ and $E_j$.
The reason for not including the context words after $E_j$ is that there are few cases in which trigger words appear after $E_j$, according to our observations.
Then, the relation extraction layer obtains the vector representations $V(C_0)$ V($E_i$), $V(C_1)$ and V($E_j$) using the BERT model and concatenates $V(C_0)$ V($E_i$), $V(C_1)$ and V($E_j$)  for relation classifier.
Our model chooses the category with the highest probability.
Finally, it produces the extracted test-oriented entities and the relations for each test case.

\subsection{Test Case Dissection into Tuples}
After extracting test-oriented entities and relations, {\tool} dissects each test case into test tuples.
During dissection, {\tool} firstly finds the extracted entities belonging to \textit{Component}, and then retrieves the associated entities based on the extracted relations. 
Finally, a \textit{Component} and an associated \textit{Behavior}, an associated \textit{Prerequisite}, an associated \textit{Manner} and an associated \textit{Constraint} make up an atomic test tuple for redundancy detection. 
Please kindly note that if there are no associated entities for an entity \textit{Component}, it is marked as ``NULL''.

Taking the Test Case \#346 as an example, {\tool} firstly retrieves two \textit{Components}, ``contents of each resource directory'' and ``visit history''.
Then, for the \textit{Component} ``visit history'', {\tool} retrieves the associated \textit{Behavior} ``switch'', and the associated \textit{Manner} ``mouse''.
After that, {\tool} constructs a tuple <``visit history'', ``switch'', NULL, ``mouse'', NULL>.
Following the above process, {\tool} iterates all the entities belonging to \textit{Component}, and constructs all the tuples.
For the two test cases in the Figure \ref{fig:granularity}, Table \ref{tab:tuples} shows all the tuples after dissection.

\begin{table}[htbp]
\Huge
  \caption{Tuples dissected from test case \#346 and \#525}
  \label{tab:tuples}
\resizebox{\columnwidth}{!}{
  \begin{tabular}{ccccccc}
    \toprule
     \textbf{Test Case} &   \textbf{Tuple} & \textbf{Component} & \textbf{Behavior}& \textbf{Prerequisite}& \textbf{Manner}& \textbf{Constraint}\\   
      \midrule
      ~  &   Tuple-1  &  contents of each     &   browse    &   NULL    &   NULL &   NULL\\
   \#346   &     &  resource directory     &       &       &    &   \\
~  &   Tuple-2 &   visit history    &   switch    &   NULL    &   mouse &   NULL   \\ 
\midrule
\#525   &   Tuple-3 &   visit history    &   browse    &   NULL    &   mouse &   NULL   \\ 

    \bottomrule
  \end{tabular}
}
\end{table}

\subsection{Detecting Redundant Test Cases by Tuple Comparison}
\label{approach:redundancy_detection}
% $Tuple_1$ $=$ <$Com_1$, $Beh_1$, $Stat_1$, $Means_1$, $Con_1$> and $Tuple_j$ $=$ <$Com_j$, $Beh_j$, $Stat_j$, $Means_j$, $Con_j$>.

% \jie{comment}
%这几个标题，形式要统一一下，都用名词或者动名词。
%Redundant Test Cases Detection by Tuple Comparison
%Test Case Dissection into Tuples
%这样是不是更好？要不感觉第四个出现了tuples，但不太清楚tuples是啥
\subsubsection{\textbf{Test Tuple Pair Construction}}
After dissecting test cases into test tuples, {\tool} builds all the tuple pairs for comparison.
A tuple pair consists of two tuples dissected from different test cases.
Take two tuples dissected from two test cases as <Com\_i, Beh\_i, Pre\_i, Man\_i, Con\_i> and <Com\_j, Beh\_j, Pre\_j, Man\_j, Con\_j>. 
Then, {\tool} judges whether two tuples in each tuple pair are semantically equivalent by comparing entities belonging to five entity categories respectively.
% \jie{comment}
%我觉得tuple i， tuple j好像在后面都没有用到过，这里需要吗。 \jie{comment}
%现在第二个发现 好像都没提features这个事情了，所以相关的说法是不是也要重新调整下。
% The following introduces the comparison methods for each entity belonging to the same entity category.
% \jie{comment}
%这里我其实觉得这个tuple pair construction只是比较的一个粒度，其实并不能算作方法的一部分。感觉这个3.4就像首先定了tuple comparison的criteria。
%其实这里我也在想，这个figure3 我们的输入是一个项目的所有test cases。这个图如果输入是两个test cases呢，就是一个项目里面任意的一对test case，是不是这个tuple pair construction就可以去掉了。因为我感觉现在3.4.3似乎主要篇幅也是在说两个测试用例怎么判断是不是redundancy的。
%这个3.4就是直接比较两个测试用例是不是redundancy的。这样是不是更直观一些？这块你们再斟酌下，我也不太确定。如果你们绝对tuple pair construction是需要的，留着也行。

% Redundancy identification is to identify tuple pairs containing 5 types of identical entities after sampling of tuple pairs. According to the representation characteristics of five types of entities, we designed different comparison similarity functions respectively to judge whether each type of entity is the same. Below we describe each type of comparison method in detail:

% \jie{comment}
%你看在figure 3里面，我觉得tuple pair construction和 comparison，redundant test case identification是一个层次的，下面的具体的component comparison，behavior comparison这些应该属于再下一个层次的。
%我觉得中间这块应该叫做tuple comparison（或者别的名字）这种感觉，然后底下包括具体的几个比较的策略。
%然后对于tuple comparison应该先写个key idea，就是比较的大概的思路是啥，现在全部分开有点太散了。
%这里是否可以把有些合并，一类是动词相关的comparison，一类型是名词相关的comparison，另一类是状态相关的comparison；这样感觉就能先说用word2vec来比较这种动词相关的，然后说名词相关的存在什么共性问题，额外用了SIF来更好的比较；当然这个具体叫法可以再看看。总体感觉现在是有些太琐碎了，没有看到high level的key idea。感觉比较像是case by case的设计了一些策略，不太好。
%而且按说这部分应该是解决一个test case有多个test feature问题的，那这部分最前面是不是应该有些key idea的陈述。

\subsubsection{\textbf{Test Tuple Comparison}}
\label{sec:tuple_comparison}
% To deal with the test cases with multiple tested features, \jie{comment}
%现在第二个发现 好像都没提features这个事情了，所以相关的说法是不是也要重新调整下。
% {\tool} detects redundancy by comparing the entities at tuple-level.
% \jie{For each tuple pair, {\tool} compares the }

For a pair of two tuples, {\tool} compares the corresponding entities belonging to the same category and judges whether they are expressing the same meaning.
To alleviate the noises bring by different expressions and better capture the semantics of the entities, we apply the word embedding technique in modeling the entities and conduct the following comparison.
Furthermore, we observe that different categories of test-oriented entities might involve different expression ways, e.g., the \textit{Component} category is usually expressed with the noun phrases as the main linguistic elements, which could influence the comparison accuracy.
To tackle this, we design three strategies for the tuple comparison, respectively for three types of expression ways corresponding with the entity categories.
% \yuan{comment}
%strategy 3中的状语从句形式的实体也算是一种词性类型吗

\begin{itemize}
\item Strategy 1, for entities in form of the verb, represent entities with word embedding and compare with cosine similarity. This strategy is applied for entity category \textit{Behavior}. 

\item Strategy 2, for entities expressed with noun phrases as the main linguistic elements, besides Strategy 1, applies the SIF method in representation. This strategy is applied for entity category \textit{Component, Manner, Constraint}. It can alleviate the noise brought by the modifier around the core noun. 

% For the entities of noun phrase, i.e., entity category \textit{component, manner, constraint}, besides \textit{Strategy 1}, we additionally employ SIF method [] to alleviate the noise bring by the modifier around the core noun. 

\item Strategy 3, for entities in form of the adverbial clause, besides Strategy 1, retrieve the indicative word for separate comparison then apply Strategy 1. This strategy is for entity category \textit{Prerequisite}. It can better distinguish the meaning of \textit{Prerequisites} for the test case.
% \yuan{comment}
% %这里怎么突然说起constraint了，应该是preresuite对应strategy 3

% , i.e., entity category \textit{prerequisite}, besides \textit{Strategy 1}, we additionally retrieve the indicative word for better distinguish the meaning.  

\end{itemize}

% \jie{delete this para.}
% For each tuple pair, {\tool} judges whether entity pair belonging to the same category is equivalent.
% % Intuitively, the naive judgment method is to check whether the corresponding strings are exactly same.
% % However, the semantically equivalent entities may have the different expressions due to the variability of natural language.
% % Therefore, {\tool} designs different semantic comparison methods for the five entity categories.}
% Intuitively, the naive judgment method is to check whether the corresponding strings are exactly same.
% However, the string comparison may fail because some the semantically equivalent entities may have the different expressions due to the variability of natural language.
% To capture the semantic representation of each entity, {\tool} firstly trains a Word2Vec model \cite{2013Efficient} using training data (details in Section \ref{experiment:dataset}).
% According to the different expression ways, {\tool} designs three comparison methods by refining the Word2Vec model: (1) the comparison for verbs, which involves the entity category \textit{Behavior}, using \textit{Word2Vec+Cos} method; (2) the comparison for entities with noun phrases as the main linguistic element, which involves the entity categories \textit{Components}, \textit{Manner} and \textit{Constraint}, using \textit{Word2Vec+SIF+Cos} method; and (3) the comparison for adverbial clause, which involves \textit{Prerequisite}, using \textit{Indicative Word Matching+Word2Vec+Cos} method.
% Following introduces the three comparison methods.

\textbf{\textit{Strategy 1: Word Embedding + Cosine Similarity.}}
Considering that the entities belonging to \textit{Behavior} are typically described as verbs, such as ``browser'' and ``visit'', and the semantic information could be accurately captured only with the word embedding technique.
We train Word2Vec model \cite{2013Efficient} using the training data (see Section \ref{extraction_baseline}), and the entity
Beh\_i and Beh\_j are vectored as $W2V_{Beh_i}$ and $W2V_{Beh_j}$ respectively using the Equation \ref{equation:w2v}:
\begin{equation}
\label{equation:w2v}
    W2V_{E} = Average(W2V_{w_1}, ...,W2V_{w_i}, ..., W2V_{w_n})
\end{equation}
where $w_i$ is the word in the extracted entity, and $W2V_{w_i}$ is the vector representation for $w_i$ returned by the trained Word2Vec model.
Then {\tool} directly calculates Cosine similarity score \cite{rahutomo2012semantic} between $W2V_{Beh_i}$ and $W2V_{Beh_j}$.
After that, Beh\_i and Beh\_j are considered semantically equivalent if the similarity score is larger than the pre-defined threshold\footnote{To ensure the high precision, the similarity threshold is set as 0.95 in {\tool}.}.

\textbf{\textit{Strategy 2: Word Embedding + SIF + Cosine Similarity.}}
For entities belonging to \textit{Component}, \textit{Manner}, and \textit{Constraint}, they are typically expressed with noun phrases as the main linguistic elements, and there are usually less informative words bringing the noise to the comparison.
For example, there are two semantically equivalent \textit{Components} ``browser application'' and ``browser'' described in two test cases.
However, the vector representations of the two entities differ a lot due to the general word ``application''.
To alleviate the noise, {\tool} additionally adopts SIF method \cite{arora2017simple} to filter out the noisy information.
It removes the projection of the average of semantic representations of an entity along semantically meaningless directions and has been proven to effectively filter the meaningless information introduced by the general words in short text \cite{hadifar2019self}.
Similarly, {\tool} judges whether two entities are semantically equivalent by the Cosine similarity and the pre-defined threshold. 
% \jie{comment}
%which has proven是在哪里proven的，要不要加个参考文献。而且如果用参考文献，就不能说in an entity了，in short text是不是更合适？
% Considering that \textit{Constraint} are usually written in [Preposition + Noun Phrase] (e,g., ``not less than 50MB/s'' and ``in Arial'', the noise also occur in \textit{Constraints} like \textit{Components} and \textit{Manners}.
% Similarly, {\tool} obtains the semantic representations, $W2V_{Con_i}$ and $W2V_{Con_j}$, use the SIF method to filter the noise and calculates the Cosine similarity.

\textbf{\textit{Strategy 3: Indicative Word Comparison + Word Embedding + Cosine Similarity.}}
% \rev{The comparison methods for adverbial clause.}
% The comparison strategy is for entities belonging to 
The entities belonging to \textit{Prerequisite} category are typically long entities described as adverbial clauses.
We observe that there are indicative words in these entities, e.g., the words determining the temporal information.
These words differ a little, yet can lead to an entirely different meaning.
For example, in two similar test cases ``Testing the CPU utilization \textit{when no preset applications are installed on the system}'' and ``Testing the CPU utilization \textit{when preset applications are installed on the system}'', the logic indicative word ``no'' in the entity \textit{Prerequisite} indicates they are different test cases. 
Another example is for the two similar test cases ``Testing hard disk can be partitioned \textit{before the system installation}'' and ``Testing hard disk can be partitioned \textit{after the system installation}'', where the temporal indicative words ``before'' and ``after'' indicate the difference. 

Taken in this sense, we summarize two lists of indicative words, i.e., words indicating \textit{logic} difference and words indicating \textit{temporal} difference. 
The example logic indicative words are ``no'', ``not'', and ``without'', while the temporal indicative words are ``after'', ``before'', ``when/while''\footnote{The full list of indicative words are displayed in our public package.}.
For the comparison, {\tool} first extracts the indicative words from the category \textit{Prerequisite}, and if they are different, the corresponding \textit{Prerequisite} are considered as non-consistent.
Otherwise, {\tool} applies Word2Vec and Cosine similarity as Strategy 1 for the comparison.

\subsubsection{\textbf{Redundancy Detection by a Tuple Covering Rule}}
\label{sec:tuple_comparison}
% Two tuples are considered semantically equivalent only if five similarity scores are all larger than the pre-defined threshold\footnote{To ensure the high precision, the similarity threshold is \jie{set} as 0.95 in {\tool}.}.
% \jie{comment}
%这里有个细节，前一个步骤是得到两个entity之间是否相同，是否不同？还是说前一个步骤得到的是两个entity 的 cosine值？
%我建议得到两个entity是否相同，因为在strategy 3，其实如果indicative words不同，就直接不同了，不需要得到sim值了。
%所以可以把这句类似的话，加到strategy 1里面去，就说这个步骤得到的是两个entity是否相同，用0.95作为阈值这种。然后这里步骤，就只是看5类entity的情况，是否都不同这样子。这样是不是逻辑更清晰。
After that, {\tool} judges whether a test case pair (TC\_i and TC\_j) is redundant according to a \textit{Tuple Covering Rule}: if the tuples for TC\_i could semantically cover tuples for TC\_j (there is a semantically equivalent tuple in TC\_i for every tuple for TC\_j), TC\_j is considered as a redundant test case.
% \jie{comment}
%有个问题我不太确定，这里只说了cover，对于两个完全一样的，算cover吗？这里要修改下说法吗
%而且我觉得这里写法需要稍微改下。第一句说两个tuple semantically equivalent是怎么样子的。然后后面一句话又在说semantically cover，里面用的是tuple set是子集，这里是不是要稍微说下semantically equivalent。就感觉说了semantically equivalent，后面就没有再说这个术语了。
% Taking the two test cases in Table \ref{tab:tuples} as an example, \#525 is considered as a redundant test case by {\tool} because tuples for \#346 could cover the tuples for \#525 (Tuple-5, Tuple-6 and Tuple-7 are equivalent to Tuple-1, Tuple-2 and Tuple-3 respectively), according to the \textit{Tuple Covering Rule}.
Using the above rule, {\tool} iterates all the test tuple pairs and detects all the redundant test cases. Please note that, for TC\_i and TC\_j, if there are the same number of tuples in the two test cases and \textit{Tuple Covering Rule} is satisfied, {\tool} considers that they are totally equivalent, and either one could be reconsidered as redundant.

\section{experiment}
\label{sec:experiment}

% This section describes research questions,  dataset, designed experiments to be conducted, baseline approaches and evaluation metrics.

\subsection{Research Questions}
\label{sec:rq}
% The research questions are formulated as following:

\textbf{RQ1: Can {\tool} effectively extract entities and relations from the test case descriptions?}
This research question aims at evaluating the effectiveness of the {\tool} in extracting five entity categories and four relation categories from test case descriptions.
    
\textbf{RQ2: Can {\tool} effectively detect the redundant test cases?}
This research question aims at evaluating the effectiveness of {\tool} in detecting redundant test cases.
% Specifically, we compare {\tool} with state-of-the-art approaches for detecting the redundant test cases.

\textbf{RQ3: How effective is each entity category for redundancy detection?}
This research question intends to investigate the performance differences of redundancy detection when removing each entity category from {\tool}.

\subsection{Subject and Dataset}
\label{experiment:dataset}
The dataset comes from our industrial partner, which is a certified third-party testing agency for software testing for over ten years.
% After developing the software, the developers will submit the software to {\company} together with the requirements specification in natural language.
% Then, \textit{system-level testing} is conduct in {\company} to demonstrate the software is compliant with the requirements.
% In the system-level testing, test engineers will analyze the requirements in the specification and manually write the NL test cases.
% Then, the test engineers will execute each test case in the manual way.
% The found defects during the testing will be reported to software developers, and regression testing is conduct after defect-fixing.
% Finally, the software will be delivered to end users if all the requirements are correctly verified.
For each software system, our industrial partner maintains a test case library.
After each test, test cases for the system will be included in the corresponding test case library.
As the system evolves, it produces redundant test cases in the test case library.
% When the new version of a software is developed, test engineers in {\company} will identified changed requirements, designs new test cases for them.
% For the unchanged requirements, test engineers typically re-execute all the test cases for the requirements in the test suite.
% The redu
% According to our industrial partner, there are no less than 20\% redundant test cases in average of all the test suites.
% \jie{comment}
%这句话要说明什么呢？no less than 20%
% Over the software evolves, the test suite results in redundancy which increase the cost of testing and maintenance.
% \sout{Our approach is to eliminate the redundant test cases in {\company}.}

In this study, we collect 3,467 test cases (TCs) from ten systems.
We retrieve the textual descriptions of the test case's summary for the redundant detection, and the average terms of each test case are also in Table \ref{tab:dataset}.
% \jie{comment}
%我个人觉得这段后面可以去掉了。我感觉没有必要写这么细，而且写那么细容易被质疑。其实就简单写下我们抽取了summary用揪心改了。
% A test case consists of the following three attributes: (1) \textit{Summary:} the general descriptions of the test case; (2) \textit{Steps:} the testing steps to be executed; and (3) \textit{Expected Behavior:} the expected behavior after executing the testing steps.
% \jie{comment}
% %没太明白这个three attributes是要怎么用，是说合并到一起抽这些内容，还是只从summary里面抽。我觉得如果只是从summary里面抽，可能需要一两句话说下为啥要这样做。
% \begin{itemize}
%     \item 
%     \textit{Summary:} the general descriptions of the test case;
%     \item
%     \textit{Steps:} the testing steps to be executed;
%     \item
%     \textit{Expected Behavior:} the expected behavior after executing the testing steps.
% \end{itemize}
% All the three attributes are written in natural language, and considered as the descriptions of a test case.
In our study, we only use the summary for redundancy detection since the summary almost covers all the target entities based on our observations.
In addition, compared with the summary, there is much noisy information in steps and expected behavior, such as testing tool installation steps and configuration steps of the testing environment.

Then, guided by our industrial partner, we iterate the test case pair in the 3,467 test cases and label the redundancy by comparing the test cases in each pair.
To guarantee the correctness of the labeling results, a labeling team with one senior researcher, one test engineer in the industrial partner and two Ph.D. students jointly work in this process. 
The redundancy is labeled according to the whole test case descriptions including summary, steps, and expected behavior.
% \jie{comment}
%这里要不要说一下，他们是根据summary，steps，expected behavior等完整的信息，来判断是不是redundancy，这能保证他们用的信息是全的，判断是准确的？
During the labeling process, each test case is labeled by one member and inspected by the other three members of the labeling team.
Once different labeling opinions arise, the final result is determined based on a team discussion and a majority voting mechanism.
The manually labeled results are considered as the ground truth set.
Table \ref{tab:dataset} shows the details of the redundancy labeling results.

\begin{table}[htbp]
  \caption{The details of the dataset
  }
  \label{tab:dataset}
\resizebox{0.9\columnwidth}{!}{
\begin{tabular}{cccc}
\toprule
\textbf{Project ID} &   \textbf{\# TCs}    &   \textbf{\# Redundant TCs} &   \textbf{Average terms}  \\   
\midrule

Project-1 &      347   &   140 &   109  \\
Project-2 &      332   &   157 &   205  \\
Project-3 &      340   &   99  &   120 \\
Project-4 &      350   &   165 &   204  \\
Project-5 &      340   &   110 &   123  \\
Project-6 &      345   &   78  &   165 \\
Project-7 &      334   &   114 &   171  \\
Project-8 &      346   &   124 &   236  \\
Project-9 &      408   &   159 &   276  \\
Project-10 &     325   &   141 &   202  \\
\midrule
\textbf{Total}          & 3,467      &  1,287 &  -     \\
\bottomrule
\end{tabular}
}
\end{table}

To train and evaluate the extraction model, we select 1,170 test cases using Stratified Sampling from 3,467 test cases according to the projects belonging to. 
For each test case, we manually label the entities and relations in the descriptions following the same process for redundancy labeling.
Finally, we labeled 2,717 entities and 1,426 relations.
In detail, for the entities, we label 1,377 \textit{Component}, 824 \textit{Behavior}, 102 \textit{Prerequisite}, 137 \textit{Manner} and 277 \textit{Constraint}.
For the relations, there are 865 \textit{Act}, 113 \textit{Require}, 145 \textit{Use} and 303 \textit{Satisfy}.

\subsection{Experiment Design}

To answer the RQ1, we used the 1,170 test cases with entity and relation labels to train and evaluate the joint extraction model.
Specifically, we adopt the randomly sampling strategy to divide the test into the training set and testing set in the ratio of 8:2.
We train {\tool} using the training set, and extract entities and relations for the testing set.
Finally, the extracted entities and relations are compared with the ground truth, and the performance is evaluated.
To avoid randomness, the above experiment is repeated five times, and the average performance is considered the final performance.
Moreover, we compare with the state-of-the-art extraction approaches, SLM and BLM (illustrated in Section \ref{extraction_baseline}).
Mann-Whitney test is used to test whether {\tool} could significantly outperform the baselines.

To answer RQ2,  we train the extraction model using all the 1,170 test cases whose entities and relations are labeled.
Then, we apply {\tool} with the trained model to the remaining 2,297 test cases, and the performance metrics are calculated by comparing the detected redundancies with the ground truth.
At the same time, we include two state-of-the-art redundancy detection approaches, CTC, Clustep, and four learning-based classifiers (illustrated in Section \ref{redundancy_baseline}), as the baselines.
Mann-Whitney test is used to test whether {\tool} could significantly outperform the baseline approaches.
% \jie{comment}
%RQ1有用test来看是否significant吗
% Second, we evaluate the non-redundant detection performance of {\tool} on the remaining 2,297 test cases.
% The performance metrics are calculated by comparing the detected non-redundancies with the ground truth.
%First, we evaluate {\tool} on the 1,170 test cases whose entities and relations are manually labeled.
%During the five experiments for RQ1, {\tool} also detects the redundancy on the testing set.
%The performance is obtained by comparing detected redundant test cases with the ground truth.
%At the same time, we include two state-of-the-art redundancy detection approach, CTC and Clustep (illustrated in Section \ref{redundancy_baseline}), as the baselines.
%
%Second, we evaluate {\tool} on the remaining 2,297 test cases whose entities and relations are not labeled.
%Specifically, We train the {\extraction} using all the 1,170 labeled test cases.
%Then, we apply {\tool} with the trained model to the remaining 2,297 test cases.
%The performance metrics are calculated by comparing the detected redundancies with the ground truth.

% To answer RQ2, we train the joint extraction model using the 1,170 test cases whose entities and relations are labeled, and evaluate {\tool} on the remaining 2,297 test cases whose entities and relations are not labeled.
% At the same time, we include two state-of-the-art redundancy detection approach, CTC and Clustep (illustrated in Section \ref{redundancy_baseline}), as the baselines.
% Mann-Whitney Test is used to test whether {\tool} could significantly outperform the baseline approaches.

To answer RQ3, we investigate the effectiveness of five category entities by ablation experiment.
We conduct five groups of experiments in the terms of ${\tool}-X$, where $X$ is each of the five entity categories.
% \jie{comment}
%这块信息量太少了，就一两句话说一下就行了，就是TSM-X 就是从中去掉了X吧，X可以是哪些这样子。
% \begin{itemize}
%     \item 
%     {\tool} - \textit{Com}: {\tool} remove the \textit{component comparison} from {\tool}, two tuples are considered as equivalent if entities belonging to other four categories are equivalent;
%     \item 
%     {\tool} - \textit{Beh}: {\tool} remove the \textit{behavior comparison} from {\tool}, two tuples are considered as equivalent if entities belonging to other four categories are equivalent;
%     \item 
%     {\tool} - \textit{Stat}: {\tool} remove the \textit{state comparison} from {\tool}, two tuples are considered as equivalent if entities belonging to other four categories are equivalent;
%     \item 
%     {\tool} - \textit{Mea}: {\tool} remove the \textit{means comparison} from {\tool}, two tuples are considered as equivalent if entities belonging to other four categories are equivalent;
%     \item 
%     {\tool} - \textit{Stan}: {\tool} remove the \textit{standard comparison} from {\tool}, two tuples are considered as equivalent if entities belonging to other four categories are equivalent.
% \end{itemize}
For each group of experiments, we train {\tool} on the 1,170 test cases and evaluate the performance of redundancy detection using 2,297 test cases.
% Mann-Whitney Test is used to test whether there are significant differences between {\tool} and each group.

\subsection{Baselines}
% There are two categories of approaches included as the baselines, i.e., entity and relation extraction baselines and redundancy detection baselines.

\subsubsection{\textbf{Entity and Relation Extraction Baselines}}
\label{extraction_baseline}
\
\newline
 \indent 
\textbf{Span-Level Model (SLM) \cite{dixit2019span}}: This is a state-of-the-art approaches for jointly extracting entities and relations.
% \rev{Similar to our extraction model, SLM also follow the joint extraction framework.}
It first obtains the candidate entities using the span strategy, then classifies the category for each entity and the relation among each entity pair.
% Unlike the previous studies which take the entity extraction and relation extraction as two individual tasks, SLM adopts a joint extraction strategy.
By combining the entity extraction loss and relation extraction loss in the training phase, SLM could avoid the error accumulation problem and outperform the approaches which individually solve the entity extraction and relation extraction tasks \cite{zheng2016joint,zheng2017joint}.  
% {\extraction} is construct by elaborating SLM.
In our study, we implement the approach strictly following its steps.
% with a help of a deep learning library pytorch \footnote{https://pytorch.org/}. 

\indent 
\textbf{BIO-Level Model (BLM) \cite{bekoulis2018joint}}: 
This is another state-of-the-art method for extracting entities and relations.
Different from SLM which takes the entity and relation extraction as a classification task, BLM models the extraction as a sequence tagging task \cite{Settles2008An}, and leverages the deep learning model to predict the label for each token.
The predicted label indicates the position of the token relative to an entity, i.e., the beginning of the entity, the inside of the entity, the end of the entity, or the outside of the entity.
After entity extraction, BLM uses a classifier to predict the relation between each entity pair.
In our study, we reuse the package provided by the paper\footnote{https://github.com/bekou/multihead\_joint\_entity\_relation\_extraction}.

\subsubsection{\textbf{Redundancy Detection Baselines}}
\label{redundancy_baseline}
\
\newline
\indent 
\textbf{Clustering Test case (CTC) \cite{tahvili2019automated}}: This is a state-of-the-art approach for detecting redundant test cases in NL. 
CTC uses the Doc2Vec algorithm \cite{le2014distributed} to generate embeddings of test cases and then groups them using two clustering algorithms HDBSCAN and FCM.
To reproduce it, following the paper, we use the gensim libarary\footnote{https://radimrehurek.com/gensim/} to build a Doc2Vec model, train the model with the test cases in the training dataset, and implement the clustering algorithms with sklearn library.
% Finally, test cases belonging to the same cluster are considered as redundancy.
 
\textbf{Clustep \cite{li2020clustering}}: This is the state-of-the-art method for detecting redundant steps in NL test cases, which is similar to our scenario since both of them involve the detection of similar test descriptions. 
Clustep encodes the descriptions by the Word2Vec model, calculates the distance between text vectors according to the relaxed word mover's distance model \cite{kusner2015word}, and detects redundant test steps using the Agglomerative clustering and k-means clustering algorithm.
We implement the approach strictly following its steps.
% \rev{This is the state-of-the-art method for detecting test redundancy among the NL test cases.
% Aiming at reducing the effort for creating scripts, 
% Considering that Clustep is also oriented towards redundancy detection for NL test cases, we also include it as the baseline.
% In out study, we implement Clustep following the steps in its study.}

% \jie{comment}.
%redundant steps 和 我们做的 redundant detection是一个事情吗

% In our study, we implement Clustep followinged steps in its study: (1) encode the description using a Word2Vec model trained on test case descriptions with the help of gensim; (2) calculate the distance between text vectors according to the relaxed word mover's distance model; (3) cluster the distance matrix using the Hierarchical Agglomerative clustering algorithm using sklearn; and (4) post-refine the clusters using k-means clustering algorithm using scikit-learn library.
% \jie{comment}
%这里要先说，这个方法的key idea是啥。然后说下是怎么复现的，怎么复现其实就简单说下，follow 他原来的研究就行。因为你看这个SLM基线，也就是说follow他们的steps 来复现的，我感觉没必要特别仔细的说用了啥。

\textbf{Learning-based Classifiers. } 
% Besides the above two primary baselines, we additionally introduce five classifier as baselines to provide more comprehensive perspectives of comparison, i.e., Naive Bayesian (NB) \cite{mccallum1998comparison}, Support Vector Machine (SVM) \cite{vapnik1999nature}, Decision Tree (DT) \cite{brijain2014survey}, TextCNN and TextRNN. 
% \jie{revised this para.}
We additionally employ machine-/deep- learning classifiers, which are commonly used in information retrieval, natural language processing and software engineering \cite{DBLP:conf/kbse/JiangTK13,DBLP:books/mk/HanKP2011,Huang2018Automating}, to predict whether two test cases are redundant. 
Its basic idea is to vectorize the test cases using the pre-trained BERT model, concatenate the vectors of each test case pair, and conduct the prediction based on it.
To provide more comprehensive perspectives of comparison, we experiment with Support Vector Machine (SVM) \cite{vapnik1999nature} ,Random Forest (RF) \cite{le2014distributed} , Decision Tree (DT) \cite{brijain2014survey} and a deep-learning based classifier TextCNN \cite{chen2015convolutional}.
The implementation is with scikit-learn library and open-source library of TextCNN\footnote{https://github.com/hellonlp/classifier\_multi\_label\_textcnn}.

\subsection{Evaluation Metrics}
We evaluate the performance from two aspects, i.e., entity and relation extraction and redundancy detection, using the four evaluation metrics, i.e., \textit{Precision, Recall, F1, and Accuracy}. 
(1) \textit{Precision}, which refers to the ratio of the number of correct
predictions to the total number of predictions; 
(2) \textit{Recall}, which refers to the ratio of the number of correct predictions to the total number of samples in the ground truth; 
(3) \textit{F1-Score}, which is the harmonic mean of precision and
recall;
(4) \textit{Accuracy} is the proportion of test cases that are correctly predicted among all test cases. 
% It is used as the overall evaluation metric.
Please note that, for two totally-equivalent test cases (details in Section \ref{sec:tuple_comparison}), either considered redundant is treated as a correct prediction.

% \subsubsection{Evaluation Metrics for Entity and Relation Extraction}
% We use three commonly used metrics to evaluate the performance of entity and relation extraction, i.e., \textit{Precision, Recall, F1}. 
% (1) \textit{Precision}, which refers to the ratio of the number of correctly extracted entities (or relations) to the total number of predictions; 
% (2) \textit{Recall}, which refers to the ratio of the number of correctly extracted entities (or relations) to the total number of entities (or relations) in ground truth set; and (3) \textit{F1-Score}, which is the harmonic mean of Precision and recall.

% \subsubsection{Evaluation Metrics for Redundancy Detection}
% We use three metrics to evaluate redundancy detection performance, i.e., \textit{ R\_Precision}, \textit{R\_Recall}, \textit{R\_F1}.
% (1) \textit{R\_Precision}, which refers to the ratio of the number of correctly predicted redundant test cases to the total number of predicted redundant test cases; 
% (2) \textit{R\_Recall}, which refers to the ratio of the number of correctly predicted redundant test cases to the total number of redundant test cases in the ground truth set; and (3) \textit{R\_F1-Score}, which is the harmonic mean of R\_Precision and R\_Recall.

% \jie{comment}
%这里感觉不是特别有必要用R-Precision这种说法吧？
%我觉得这块可以不用分成两个subsection，就直接说有两个方面需要评估，一个是entity and relation 抽取的准不准，二是redundancy 检测的准不准，这两个任务我们都用常用的Precision，recall和F-measure。然后那redundancy举例，这三个指标是怎么算的。

\section{results and analysis}
\label{sec:result}
% This section presents the evaluation results in order to answer the research questions. 

% \subsection{Redundancy Detection (RQ2)}
% \label{results:rq2}

\subsection{Entity and Relation Extraction Performance}

Table \ref{tab:extraction_performance} shows the performances of entity and relation extraction of our extraction model and baselines . 
In general, the extraction model could achieve promising performance with 97.5\% precision, 94.8\% recall for entity extraction, and 90.4\% precision, 97.6\% recall for relation extraction.

Compared with the baseline, SLM, our extraction model is 4.5\%, 9.5\% higher in F1 for entity and relation respectively.
The results indicate the effectiveness of introducing global information and context information into the traditional extraction models.
There is only one exception appearing in the recall of \textit{Prerequisite}.
It is that the entities belonging \textit{Prerequisite} are typically described as adverbial clauses whose length is much larger than the other four categories.
It is difficult for BERT to encode the long entities.
Among the three models, BLM shows the worst performance, which indicates the advantage of the span-based joint extraction approach for test case descriptions.
In addition, for each project, the average time to train the Tscope model is 72 seconds, yet it can be done offline.

 \begin{table}[htbp]
  \caption{Performance of entity and relation extraction}
  \label{tab:extraction_performance}
\resizebox{\columnwidth}{!}{
\begin{tabular}{cccccccccccc}
\toprule
\multirow{2}{*}{\textbf{Metric}} &
  \multirow{2}{*}{\textbf{Model}} &
  \multicolumn{5}{c}{\textbf{Entity Categories}} &
  \multicolumn{5}{c}{\textbf{Relation Categories}} \\ \cline{3-7} \cline{9-12} 
 &
   &
  \multicolumn{1}{c}{\textit{\textbf{Com}}} &
  \multicolumn{1}{c}{\textit{\textbf{Beh}}} &
  \multicolumn{1}{c}{\textit{\textbf{Pre}}} &
  \multicolumn{1}{c}{\textit{\textbf{Man}}} &
  \textit{\textbf{Con}} &
\textbf{ } &
  \multicolumn{1}{c}{\textit{\textbf{Act}}} &
  \multicolumn{1}{c}{\textit{\textbf{Require}}} &
  \multicolumn{1}{c}{\textit{\textbf{Use}}} &
  \textit{\textbf{Satisfy}} \\ \midrule
\multirow{3}{*}{{Precision}} &
  \textit{\extraction} &
  \multicolumn{1}{c}{\cellcolor{pink}\textbf{99.1\%}} &
  \multicolumn{1}{c}{\cellcolor{pink}\textbf{99.2\%}} &
  \multicolumn{1}{c}{\cellcolor{pink}\textbf{94.4\%}} &
  \multicolumn{1}{c}{\cellcolor{pink}\textbf{97.3\%}} &
  \cellcolor{pink}\textbf{97.8\%} &
   &
  \multicolumn{1}{c}{\cellcolor{pink}\textbf{90.3\%}} &
  \multicolumn{1}{c}{\cellcolor{pink}\textbf{91.2\%}} &
  \multicolumn{1}{c}{\cellcolor{pink}\textbf{90.1\%}} &
  \cellcolor{pink}\textbf{90.2\%} \\ 
 &
  \textit{SLM} &
  \multicolumn{1}{c}{98.8\%} &
  \multicolumn{1}{c}{95.2\%} &
  \multicolumn{1}{c}{75.1\%} &
  \multicolumn{1}{c}{91.8\%} &
  94.9\% &
   &
  \multicolumn{1}{c}{72.8\%} &
  \multicolumn{1}{c}{69.2\%} &
  \multicolumn{1}{c}{85.3\%} &
  78.1\% \\ 
 &
  \textit{BLM} &
  \multicolumn{1}{c}{91.2\%} &
  \multicolumn{1}{c}{88.7\%} &
  \multicolumn{1}{c}{73.9\%} &
  \multicolumn{1}{c}{83.1\%} &
  85.2\% &
   &
  \multicolumn{1}{c}{84.1\%} &
  \multicolumn{1}{c}{69.8\%} &
  \multicolumn{1}{c}{\cellcolor{pink}\textbf{90.1\%}} &
  81.2\% \\ \midrule
\multirow{3}{*}{Recall} &
  \textit{\extraction} &
  \multicolumn{1}{c}{\cellcolor{pink}\textbf{95.2\%}} &
  \multicolumn{1}{c}{\cellcolor{pink}\textbf{96.9\%}} &
  \multicolumn{1}{c}{93.2\%} &
  \multicolumn{1}{c}{\cellcolor{pink}\textbf{95.1\%}} &
  \cellcolor{pink}\textbf{93.8\%} &
   &
  \multicolumn{1}{c}{\cellcolor{pink}\textbf{97.9\%}} &
  \multicolumn{1}{c}{\cellcolor{pink}\textbf{99.3\%}} &
  \multicolumn{1}{c}{\cellcolor{pink}\textbf{97.2\%}} &
  \cellcolor{pink}\textbf{96.1\%} \\  
 &
  \textit{SLM} &
  \multicolumn{1}{c}{93.2\%} &
  \multicolumn{1}{c}{92.1\%} &
  \multicolumn{1}{c}{\cellcolor{pink}\textbf{98.9\%}} &
  \multicolumn{1}{c}{93.2\%} &
 88.1\%  &
  &
  \multicolumn{1}{c}{94.8\%} &
  \multicolumn{1}{c}{98.8\%} &
  \multicolumn{1}{c}{95.1\%} &
  96.0\% \\  
 &
  \textit{BLM} &
  \multicolumn{1}{c}{94.2\%} &
  \multicolumn{1}{c}{90.9\%} &
  \multicolumn{1}{c}{82.2\%} &
  \multicolumn{1}{c}{84.8\%} &
  90.1\% &
   &
  \multicolumn{1}{c}{77.2\%} &
  \multicolumn{1}{c}{67.1\%} &
  \multicolumn{1}{c}{65.9\%} &
  76.2\% \\ \midrule
\multirow{3}{*}{F1} &
  \textit{\extraction} &
  \multicolumn{1}{c}{\cellcolor{pink}\textbf{97.1\%}} &
  \multicolumn{1}{c}{\cellcolor{pink}\textbf{98.0\%}} &
  \multicolumn{1}{c}{\cellcolor{pink}\textbf{93.8\%}} &
  \multicolumn{1}{c}{\cellcolor{pink}\textbf{96.2\%}} &
  \cellcolor{pink}\textbf{95.7\%} &
   &
  \multicolumn{1}{c}{\cellcolor{pink}\textbf{93.9\%}} &
  \multicolumn{1}{c}{\cellcolor{pink}\textbf{95.1\%}} &
  \multicolumn{1}{c}{\cellcolor{pink}\textbf{93.5\%}} &
 \cellcolor{pink}\textbf{93.1\%}  \\  
 &
  \textit{SLM} &
  \multicolumn{1}{c}{95.9\%} &
  \multicolumn{1}{c}{93.6\%} &
  \multicolumn{1}{c}{85.3\%} &
  \multicolumn{1}{c}{92.5\%} &
  91.3\% &
   &
  \multicolumn{1}{c}{82.3\%} &
  \multicolumn{1}{c}{81.3\%} &
  \multicolumn{1}{c}{87.9\%} &
  86.1\% \\  
 &
  \textit{BLM} &
  \multicolumn{1}{c}{92.6\%} &
  \multicolumn{1}{c}{89.7\%} &
  \multicolumn{1}{c}{77.8\%} &
  \multicolumn{1}{c}{83.9\%} &
  87.5\% &
   &
  \multicolumn{1}{c}{80.5\%} &
  \multicolumn{1}{c}{68.5\%} &
  \multicolumn{1}{c}{76.1\%} &
  78.6\% \\ \bottomrule
\end{tabular}
}
\end{table}

% \begin{figure}[htb]
%     \begin{center}
%     \includegraphics[height=5cm,width=8cm]{fig/entity_extract1.eps}
%     \end{center}
%     \caption{entity extraction result.}
%     \label{fig:entity_extract}
% \end{figure}

% \begin{figure}[htb]
%     \begin{center}
%     \includegraphics[height=5cm,width=8cm]{fig/relation_extract1.eps}
%     \end{center}
%     \caption{relation extraction result.}
%     \label{fig:relation_extract}
% \end{figure}

% \input{tab/extract_performance}

\subsection{Redundancy Detection Performance}

\begin{figure}[htbp]
\setlength{\abovecaptionskip}{1pt}
\setlength{\belowcaptionskip}{1pt}
    \begin{center}
    \includegraphics[height=5.4cm,width=8cm]{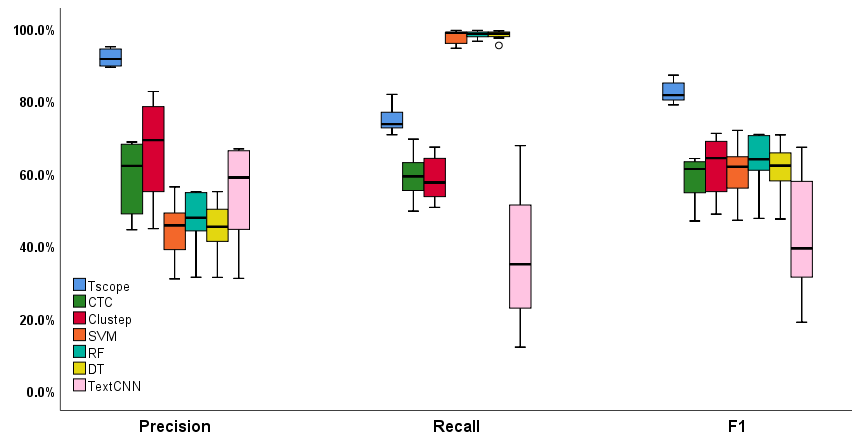}
    \end{center}
    \caption{Redundancy detection performance on ten projects}
    \label{fig:redundancy_extract}
\end{figure}

Table \ref{tab:redundancy_overview} shows the performance of detecting redundant and non-redundant test cases.
In general, {\tool} could reach 91.8\% precision and 74.2\% recall for redundant test cases, 86.7\% precision and 96.0\% recall for the non-redundant test cases.
The overall classification accuracy is 88.0\%.
Figure \ref{fig:redundancy_extract} shows the redundancy detection performance for {\tool} and baselines on the ten projects.
The results reveal that {\tool} outperforms two state-of-the-art redundancy detection approaches (CTC and Clustep), and four commonly-used classifiers (SVM, RF, DT, and TextCNN) in F1.
In addition, {\tool} shows the smallest variance of the seven approaches in precision and F1, which reflects the high robustness.

% Please add the following required packages to your document preamble:
% \usepackage{multirow}
\begin{table}[htbp]
\caption{Performance of redundancy detection}
\label{tab:redundancy_overview}
\resizebox{\columnwidth}{!}{
\begin{tabular}{ccc|c}
\toprule

\textbf{Metric} & \textbf{Redundant} & \textbf{Non-Redundant} & \textbf{Accuracy} \\ \midrule

\textit{Precision}  & 91.8\% &  86.7\% & \multirow{3}{*}{88.0\%} \\

\textit{Recall} & 74.8\% & 96.0\% & \\

\textit{F1} & 82.4\%  & 91.1\% &     \\
\bottomrule
\end{tabular}
}
\end{table}

Especially, {\tool} shows the great advantages in precision.
Compared with the two redundancy detection approaches, {\tool} is 39.4\% higher than CTC and 30.5\% higher than Clustep in precision.
Of the four commonly-used classifiers, the precision is not promising (the best precision is 66.6\% achieved by TextCNN).
The results of the Mann-Whitney test show the differences between {\tool} and each baseline approach are all significant (the significant level is 0.05).
It is mainly that the traditional approaches and classifiers can not perceive the subtle differences in the test-oriented entities.
The results emphasize the effectiveness of {\tool} for accurate redundancy detection.

% Moreover, the results of the Mann-Whitney test show the differences between {\tool} and each baseline approach are all significant (significant level is 0.05).
% There are two main reasons: (1) the previous approaches use all the words in descriptions when vectorizing, and they can not distinguish the non-redundant test cases with similar descriptions (e.g, TC \#117 and TC \#123 in Figure \ref{fig:similar_tcs}); and (2) previous approaches could not handle the test cases involving multiple tested features (e.g., TC \#346 in Figure \ref{fig:granularity}).
% In our study, we identify five entity categories for redundancy detection by the empirical analysis (see Section \ref{sec:background}.
% Unlike the previous studies, {\tool} could reach promising Precision by decomposing test cases and strictly comparing each entity category.
% The results also imply the necessity of entity extraction and test case decomposition. 

As for the recall, we notice that the commonly-used classifiers could reach 94.4\%-99.1\% in the recall.
The reason is that the classifiers tend to predict all similar test cases as redundancy.
However, it has obvious damage to the precision.
In redundancy detection, precision should be guaranteed first compared with recall, since low precision may affect the fault-revealing power of a test suite (discussed in Section \ref{sec:Precision_importance}).
Compared with CTC and Clustep, {\tool} shows no significant differences in the recall, which indicates {\tool} could retrieve almost the same magnitude of redundancy with higher precision.

There are 25.2\% (100\% - 74.8\% in recall) redundant test cases in ground truth that are not retrieved.
The main reason for true negatives is due to some test-oriented entities are missing.
For example, there are two descriptions, i.e., ``Test whether the system can display the calendar function correctly'' and ``Test whether the calendar function in the system is correct''.
The test cases have the same component ``calendar function'', while the former contains a behavior ``display'' and no obvious behavior in the latter.
The two test cases are both considered non-redundant by {\tool} due to the differences in the category \textit{Behavior}.
However, they are labeled as redundancy in the ground truth, due to ``display'' in the former does not introduce much information for distinguishing.
However, to avoid ambiguity, it would be better to clearly describe the necessary test-oriented entities in the descriptions.
It is worth noting that for each project Tscope requires an average of 42 seconds for redundant detection, compared to 15 seconds for baselines. For memory, Tscope takes about 400MB, while baselines take 200-250MB. In summary, Tscope consumes comparable cost, yet achieves significantly higher performance.

% By further analyzing the true negatives, 

% \yuan{However, according to our observations, there are two main reasons for the low recall rate:(1) Some test case descriptions are of poor quality, omitting key entities that do not prevent human understanding, but the model cannot fully extract its information, resulting in the final cumulative error. 
% }

% \yuan{In addition to the advantages of performance metrics, the fluctuation range of the F1 value of {\tool} on 10 items is lower than other baselines, which reflects the high robustness of our model.}

% \yuan{Table \ref{tab:non-redundancy_project} shows the Non-redundancy detection performance on the ten projects.
% This is the opposite of redundant detection performance and is used to reflect the balance of the data.
% The average Precision of the ten projects is 85.31\%, the average recall is 95.72\%, and the average F1 value is 90.18\%. 
% The non-redundant recall rate is higher than 95\%, which indicates that very few non-redundant test cases are incorrectly classified as redundant entities, further explaining the reason for the high-Precision values of redundant detection.}

% \input{tab/project_nonredundancy}

\subsection{Entity Category Effectiveness}
\label{sec:rq3}

Table \ref{tab:rq3} shows the performance on redundancy detection after removing different entity categories from {\tool}.
% The evaluation metrics in redundancy detection are Redundancy\_Precision(R\_Precision), Redundancy\_Recall(R\_Recall) and Redundancy\_F1(R\_F1).
% R\_Precision, R\_Recall and R\_F1 represents the evaluation metrics for redundancy detection.
The figure in the bracket is the difference compared to \textit{{\tool}}.
% The symbol ``*'' is suffixed with the figure if the Mann-Whitney Test shows the difference is significant (significant level is 0.05).
In general, the overall performances (F1) significantly decrease after removing most of the entity categories except for \textit{Prerequisite}, which indicates the necessity of the five entity categories.

In detail, the difference is greatest when removing \textit{Component}, since the \textit{Component} is the basic category and the frequency of occurrence in the descriptions is the highest among the five entity categories.
The difference is smallest after removing \textit{Prerequisite}.
Although \textit{Prerequisite} shows the greatest correlation coefficient (0.841) with the variable $Redundant$, it appears least among the five categories in our dataset.
Therefore, the difference is not obvious in F1 after removing \textit{Prerequisite}.
However, the difference is significant in R\_Precision, which also indicates the importance of accurate redundancy detection.

\begin{table}[htbp]
\caption{Performance after removing each entity category}
\label{tab:rq3}
\resizebox{\columnwidth}{!}{
\begin{tabular}{cccc}
\toprule
\textbf{Experiment Group} & \textbf{Precision} & \textbf{Recall} & \textbf{F1} \\ \midrule

{\tool} &   91.8\%    &   74.8\%    &  82.4\%  \\ 

{\tool} - \textit{Com}  & 52.2\% (-39.6\%)  & 32.7\% (-42.1\%)  &   40.2\% (-42.2\%)    \\ 

{\tool} - \textit{Beh}  &   66.6\% (-25.2\%)    & 67.0\% (-7.8\%)  &   67.2\% (-15.2\%)    \\ 

{\tool} - \textit{Pre} &   86.3\% (-5.5\%)    &   67.0\% (-7.8\%)    &   75.2\% (-7.2\%)    \\ 

{\tool} - \textit{Man}  &   84.1\% (-7.7\%)    &   65.5\% (-9.3\%)    &   73.8\% (-8.6\%)    \\ 

{\tool} - \textit{Con} &   83.3\% (-8.5\%)    &   66.4\% (-8.4\%)    &   73.8\% (-8.6\%)    \\ 

\bottomrule
\end{tabular}
}
\end{table}

\section{discussion}
\label{sec:discussion}
% This section presents the learned lessons in practice and threats to validity. 

\subsection{The Importance of High Precision for Redundancy Detection}
\label{sec:Precision_importance}
Our study focuses on redundancy detection for NL test cases, and it could be applied to many research problems, such as test suite reduction (or minimization), test case selection and test case prioritization \cite{yoo2012regression,lou2019chapter}.
In the previous studies, \textit{Precision} and \textit{Recall} are two commonly-used metrics to evaluate the performance of the proposed approaches.
It has been reported that precision and recall are two competing metrics, and they need to be balanced to achieve promising overall performance in the previous studies \cite{gordon1989recall,buckland1994relationship}.
However, we consider that precision has a higher priority compared with recall when designing automatic approaches for redundancy detection.

For example, in the test suite reduction, the detected redundant test cases will be removed from the set to be executed, aiming at reducing the testing cost \cite{hsu2009mints}.
The lower precision means that more non-redundant test cases will be removed from the executed set, which may reduce the fault-revealing power of the test suites. 
Thus, automated approaches should improve performance based on the premise of precision.
As we illustrate in Section \ref{sec:introduction}, previous studies suffer low precision since they cannot perceive the subtle differences in the test-oriented entities in test cases. 
% \jie{comment}
%我觉得这块表达后面等再改改 similar test cases are not always redundant不太确切。因为现在这种先抽实体再比较，也是找的similar。感觉应该是说简单通过文本匹配不能完全确定测试用例是否redundant这种？
It is also reported that current approaches often compromise fault detection effectiveness of a test suite by existing empirical studies \cite{rothermel2002empirical,marijan2018practical}.
To overcome the issue, {\tool} defines five test-oriented entity categories and detects the redundancy by the fine-grained comparison of the five entities.
The evaluation also shows that {\tool} could reach promising performance, especially for precision, which indicates that {\tool} could be more effective in practice.

% \subsection{The learned lessons}
% \subsubsection{The}

\subsection{Additional Benefits of the Fine-grained Redundancy Detection}

Besides achieving high accuracy, our proposed fine-grained redundancy detection approach also has the following additional benefits.

\textit{\textbf{The interpretability of redundancy detection}}.
% \rev{{\tool} could not only detect the redundant test cases but also potentially increase the interpretability of detection.}
Previous approaches for redundancy detection are typically black-box approaches since they can not perceive subtle differences in test-oriented entities.
On the contrary, {\tool} conducts the comparison for the test-oriented entities, and calculates the similarity scores of the entities belonging to the five entity categories respectively (details in Section \ref{sec:tuple_comparison}).
The entities whose similarity score is less than the pre-defined similarity threshold could be considered as the reasons for non-redundancy.
% \jie{comment}
%这里我觉得得正向写，得说redundancy的原因。因为redundancy是要去掉的，你要去掉肯定要告诉人家原因是啥？non-redundancy本来就是不管的，这里感觉不需要说原因。
%然后我觉得这几条，应该在前面或者后面加上这个事情的作用。例如可解释性带来啥好处。特别是第二点，优化执行序列有啥好处，或者是为啥要优化执行序列，这个事情为啥重要，这种感觉的。我都加上了，需要的话可以再调整下
Explicitly presenting the reasons for the redundancy can potentially increase the confidence of testers in managing the test cases.
% On the other hand, redundant test cases indicates all the testing-oriented entities could be covered by another test case.

\textit{\textbf{Optimization of the execution sequence of test cases}}.
% \rev{Our fine-grained approach may introduce additional benefits to optimize the execution sequence of test cases.}
The execution of the whole test suite usually takes a long time, and optimizing the execution sequence for reducing the time is valuable \cite{fang2014similarity}. 
The fine-grained test case information can potentially facilitate this task.
For example, if some non-redundant test cases have the equivalent entities belonging to \textit{Prerequisite}, it indicates that they share the same test prerequisite.
In practice, prerequisite preparation is cost-consuming, especially for load testing or extreme testing.   
It could be better to continuously execute test cases with the equivalent prerequisites to avoid repeatedly preparing the testing prerequisites, which potentially reduces the testing cost in practice.

\textit{\textbf{Improving the writing quality}}.
The quality of test specifications and test case descriptions is the premise to ensure test quality.
How to write complete and unambiguous test cases or test specifications is a challenging task in practice \cite{zander2008quality}.
The five test-oriented entity categories could be considered as the description items to guide the engineers to write test cases.
And our fine-grained approach, {\tool}, can be used to check the completeness of the test case descriptions or test specifications in real-time, and remind the test engineers of the missing description items to improve the writing quality.

\subsection{Other Applications of Entity Categories}
\begin{figure}[htbp]
\centering
\includegraphics[width=\columnwidth]{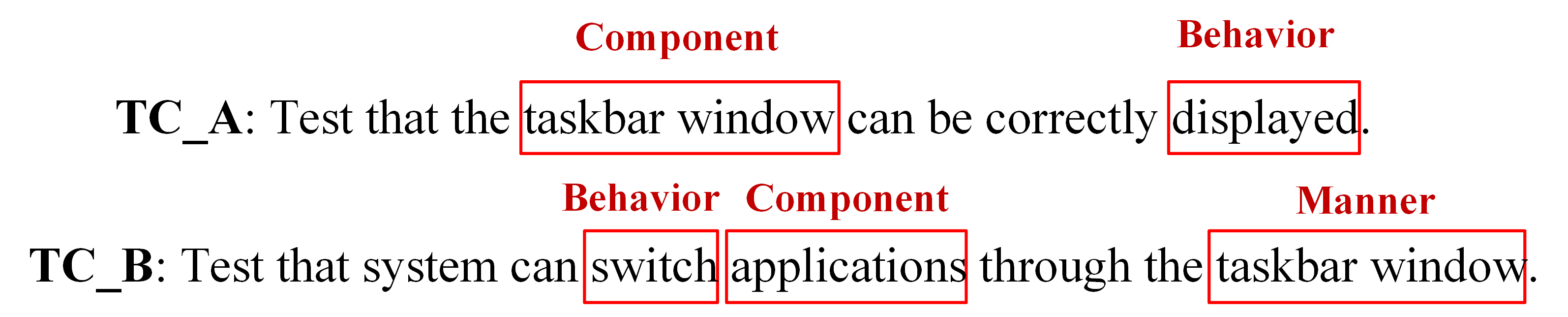}
\caption{The examples of two dependent test cases
}
\label{fig:dependent_tcs}
\end{figure}

In our study, we define five entity categories for redundancy detection.
The entity categories are not limited to redundancy detection, but other tasks, such as test dependence detection \cite{tahvili2018functional} and requirements-test linking \cite{uusitalo2008linking}.
For example, Figure \ref{fig:dependent_tcs} shows summaries of two test cases and the entities extracted by {\tool}.
According to the definition of test dependence \cite{arlt2015if,bell2014detecting}, \textit{TC\_B} is dependent on \textit{TC\_A} since if the ``taskbar window'' can not be displayed (TC\_A fails), the application can not be switched by the taskbar window (TC\_B can not succeed).
The dependence could be detected using a heuristic rule: ``TC\_B is dependent on TC\_A if \textit{Component} in TC\_A is the same as the \textit{Manner} in TC\_B''.

The entity categories could also be used to reconstruct the link between test cases and requirements.
Users could manually label \textit{Component}, \textit{Behavior}, \textit{Manner}, \textit{Prerequisite} and \textit{Constraint} from the requirements descriptions, and trains the model to extract the five entity categories.
After that, a link between a test case and a requirement can be reconstructed if the entities in each category are equivalent.

\subsection{Validity Threats}
\label{sec:validity}

% The validity threats involve external validity, internal validity and construct validity. 

\textbf{{External Validity}}.
The external threats are related to the generalization of the proposed approach.
First, we experimented with the data taken from one company.
The results may be different in other scenarios. 
However, we evaluate the performance on ten projects from different domains, which could reduce this threat. 
% Second, the test cases used in our study are in natural language, which may not be appreciated by other forms of test cases, e.g., executable scripts containing command lines and code. 
% However, {\tool} does not utilize unique characteristics except for natural language description, which could alleviate the threat. 
Second, we only use the summaries for extracting test-oriented entities and relations, which may ignore some entities buried in the other attributes such as steps and expected behavior.
However, summaries almost cover all the target entities based on our observations.
In addition, the Cohen\' kappa coefficient between $EQ_{all}$ and $Redundant$ is 0.984 (details in Section \ref{sec:correlation_analysis}).
The results imply that almost redundancy could be detected by the entities and relations in summaries, which could alleviate the threat.
% Third, our corpus is written in Chinese.
% Due to the differences in linguistic morphology, the proposed approach may not be suitable for other languages.
% While, when switching to other languages, the method only needs to be slightly adjusted in data preprocessing.
% For example, when dealing with English, the approach could work after performing some preprocessing steps such as stemming each word into its morphological root.

\textbf{{Internal Validity}}. 
The internal threats relate to experimental errors and biases.
First, the four relation categories are built around the \textit{Component}, and the other four entity categories are associated with \textit{Component}.
We find that a few cases that \textit{Prerequisite} or \textit{Constraint} is more accurately associated with the combination of a \textit{Behavior} and a \textit{Component} rather than a single \textit{Component}.
It may introduce bias when dissecting test cases and redundancy detection.
However, such cases account for a very small proportion, which could alleviate the threat.
% Second, we evaluate the performance of entity and relation extraction with a small number for labeled test cases.
% The limited number of test cases for evaluation may introduce randomness of the performance.
Second, for the baselines whose packages are not provided, it may introduce bias in the implementation processes.
However, we strictly follow the steps in their studies, which may alleviate the threat.

% \jie{comment}
%我觉得有一条threats应该说是基线的实现这种，就说我们是严格按照方法描述来实现的等等这些。

% \textbf{{Construct Validity}}. The construct threats relate to the suitability of evaluation metrics. 
% We use \textit{Precision}, \textit{Recall}, and \textit{F1} as the measurement metrics to evaluate the performance of {\tool}.
% We consider {\tool} could be effectively applied to industrial practice based on the measurement metrics.
% It would be better if we investigated the effort savings to evaluate the performance of {\tool}.
% However, the \textit{Precision}, \textit{Recall} and \textit{F1} are the commonly-used metrics in the machine learning and the information retrieval fields \cite{Berry2017Panel}.
% Moreover, we compare the {\tool} with state-of-the-art approaches and developers respectively using these metrics, which could indicate the advantages of the {\tool}.

% \textbf{{Conclusion Validity}}. The conclusion threats relate to the appropriateness of the conclusion.
% In the RQ4, we adopt the greedy strategy and empirically choose 15 combinations to tune the three bootstrapping parameters.
% The corresponding best result of the 15 combinations is considered as the final performance of {\tool}.
% However, the optimal one may not exist in the 15 parameter combinations.
% Accordingly, the results reported in RQ1 may not be the best performance that {\tool} could achieve.
% However, it is impossible to enumerate all the parameter combinations, and the adopted strategy could reach promising results.
% Thus, the threat could be alleviated.
\section{related work}
\label{sec:related work}

% \rev{Our study is related to the test redundancy detection.}
% There are test suite reduction (or minimization) techniques which aim to remove redundant test cases from the original test suite but preserve its fault detection capabilities.
% Many redundancy detection approaches in the literature are in the context of regression tests in continuous integration (CI).
% Based on the assumption that the source code is available in CI, these
% \rev{In previous studies, numerous }
% Redundancy of test cases \jie{is becoming prevalent} \rev{has been studied is a critical problem} in practice \cite{jeffrey2005test,fraser2007redundancy,zamli2018hybrid}.

Test cases act as the starting point for the test execution and the following quality assurance activities \cite{jeffrey2005test,fraser2007redundancy,zamli2018hybrid}.
Redundant test cases frequently appears in the test suite and potential affects many automatic techniques, e.g., test suite reduction \cite{rothermel2002empirical,fraser2007redundancy,cruciani2019scalable}, test case selection \cite{cartaxo2011use,yoo2012regression,bekoulis2018joint} and test case prioritization \cite{fang2012comparing,fang2014similarity,shin2018test}.

% Our study focuses on redundancy detection for NL test cases, and it could be applied to many research problems, such as test suite reduction (or minimization), test case selection and test case prioritization \cite{yoo2012regression}.

Many studies are focusing on the redundant detection of the white-box test.
Among them, several approaches employed the test coverage metrics for redundancy detection.
For example, Offutt et al. \cite{offutt1995procedures} leveraged the statement coverage to detect the redundancy in a test suite.
Rothermel et al. \cite{rothermel1998empirical} proposed an approach to detection redundancy in a test suite based on the branch coverage.
The basic assumption of the approaches is that if two test cases have the same test coverage metric, either is considered redundancy.
Koochakzadeh \cite{koochakzadeh2010tester} reported that the coverage information suffers from a large number of false-positive errors.
They combined the coverage information with additional tester-assisted information to improve the precision of redundancy detection.
Different from our study, these studies focus on the white-box test.
Yoo et al. \cite{yoo2012regression} reported that this white-box coverage with the tester-assisted information is costly or even biased.

There were also approaches applying information retrieval techniques for the white-box test case redundant detection\cite{chen2010adaptive,liu2011user,marijan2018practical,cruciani2019scalable}.
For these approaches, different similarity metrics are proposed and two test cases with the high measurement metric are considered redundancy. 
For example, 
Chen et al. \cite{chen2010adaptive} proposed an approach to exploit the diversity among test cases for guiding selection.
It first selects a random set of test cases and then filters redundancy based on their distance from the already selected test cases.
Cruciani et al. \cite{cruciani2019scalable} leveraged the vector space model to transform the test case into points in the Euclidean space, and then it detects the redundancy using the clustering algorithms. 
Different from our study, the inputs of these approaches are typically executable test cases that consist of test source code and command lines, rather than natural language.

There are studies focusing on the redundant detection of black-box tests.
Cartaxo et al. \cite{cartaxo2011use} and Hemmati et al. \cite{hemmati2013achieving} proposed approaches to reduce the model-based test suite with similarity comparison. 
However, they mainly relied on a formal model of program behavior such as LTS and UML diagrams, which are not available in some cases or are biased.

Several studies aimed at detecting redundant NL test cases.
Tahvili et al. \cite{tahvili2019automated} used an implementation of Doc2Vec algorithm to generate embeddings of test cases and then groups them using two clustering algorithms HDBSCAN and FCM.
Li et al. \cite{li2020clustering} designed an approach to identify similar test steps in the textual test cases.
It leverages the word embedding technique along with Relaxed Word Mover's Distance to analyze the similarity of test steps, then combines hierarchical and K-means clustering algorithms to detect similar test steps.
Viggiato et al. \cite{viggiato2021identifying} used a combination of text embedding, text similarity, and clustering techniques to identify similar NL test cases based.
These approaches suffer from low accuracy because they treat the test case as a whole and cannot capture the fine-grained semantic information as our approach.

% To alleviate the issue, our approach extracts the key entities and only compares the entities instead of the whole test case descriptions.
% The clustering-based approaches are included as the baseline approaches to investigate whether our approach could outperform these approaches.

\section{conclusion}
\label{sec:conclusion}
% In software testing, system-level test cases are typically written in natural language and executed manually.
Due to the redundancy of requirements, parallel testing, and tester turnover within long evolving history, there are lots of redundant test cases.
As the software evolves, redundant test cases significantly increase the cost of test and maintenance efforts.
% In the previous studies, numerous approaches have been proposed to detect and eliminate the redundant test cases.
Previous redundancy detection approaches suffer low accuracy because of their weakness in the capture of the test case's fine-grained semantic information and inherent meaning.
In this study, we re-formulate the problem and propose a fine-grained approach {\tool} to detect redundancy from test cases in natural language.
{\tool} extracts the test-oriented entities and associated relations to dissect the NL test case into atomic test tuple(s), and conduct similarity comparison on them.
% employs the fine-grained test tuples to compare the similarity.
Evaluation shows {\tool} could outperform the state-of-the-art approaches for entity and relation extraction and redundancy detection.
Moreover, the results also demonstrate the contribution of our defined five categories of entities in redundancy detection, which further indicates our problem formulation is promising.
% on the dataset shows that the entity categories in {\tool} could effectively detect the redundancy.

In the future, we will further investigate the cost-effectiveness of {\tool} in real-world practice. 
Apart from that, we will apply the fine-grained approach based on entity and relation extraction to other software engineering tasks, such as duplicate issue report detection, test dependence detection, and requirements-test linking.
% test dependence detection.
% and requirements-testing linking.

% the industrial environment.
% In addition, we will apply the test-oriented entity and relation extraction to more software engineering tasks, such as 
\section*{Acknowledgments}
This work was supported by the National Key Research and Development Program of China under Grant 2018YFB1403400, and in part by the National Natural Science Foundation of China under Grants 62072442. 

\bibliographystyle{ACM-Reference-Format}
\bibliography{reference}

\end{document}